\documentclass[fleqn,10pt]{wlscirep}

\usepackage[T1]{fontenc}
\usepackage[utf8]{inputenc}
\usepackage{babel}
\usepackage{amsmath}
\usepackage{graphicx}
\usepackage{hyperref}
\makeatletter
\usepackage{physics}

\makeatother

\begin{document}
\title{Dynamically Tunable Anisotropic Rabi Model in Circuit QED}

\author[1,*]{S. Mojtaba Tabatabaei}
\author[2]{Babak Zare Rameshti}
\author[1]{Mohsen Akbari}
\affil[1]{Department of Physics, Kharazmi University, Tehran 15719-14911, Iran}
\affil[2]{Department of Physics, Iran University of Science and Technology, Narmak, Tehran 16844, Iran}

\affil[*]{Corresponding author: S. Mojtaba Tabatabaei (s.m.taba90@gmail.com , moj.taba@khu.ac.ir)}

\begin{abstract}
We propose a circuit QED architecture that realizes the full anisotropic
Rabi model (ARM) with in-situ tunability of its interaction symmetry.
The circuit comprises a superconducting qubit coupled to a resonator
through two simultaneous channels: a direct inductive path and a flux-tunable
capacitive path mediated by a transmon coupler. By controlling the
external magnetic flux, the destructive and constructive interferences
between these paths continuously tune the ratio of the Jaynes-Cummings
(JC) and anti-Jaynes-Cummings (AJC) couplings, sweeping the system
across the entire ARM parameter space-{}-{}-from pure JC, through
the standard Rabi regime, to pure AJC-{}-{}-without requiring rotating-wave
approximations or external parametric drives. Furthermore, we identify
a coherence sweet spot where the dispersive shift and photon shot-noise
dephasing are suppressed. In the pure AJC regime, we demonstrate that
the vacuum Rabi splitting vanishes and the Purcell decay rate is strongly
mitigated while preserving high qubit readout contrast.
\end{abstract}
\maketitle

\section*{Introduction}

The ability to coherently couple artificial atoms to quantized electromagnetic
modes underlies many of the key developments in quantum electronics,
from fundamental studies of light--matter interaction to the implementation
of superconducting quantum processors. In circuit quantum electrodynamics
(circuit QED), this interaction can be engineered with unprecedented
precision by integrating superconducting qubits with on-chip microwave
resonators, enabling strong and even ultra-strong coupling regimes
that go far beyond the reach of conventional cavity QED~\cite{Blais2021,Schoelkopf2008}.
A typical platform in circuit-QED consists of a superconducting qubit
coupled either capacitively or inductively to a superconducting coplanar
waveguide resonator~\cite{Wallraff2004,Blais2004}. The underlying
light-matter interaction between the cavity mode and the qubit is
described by the Rabi model which is one of the most fundamental models
of light--matter interaction in quantum physics~\cite{Rabi1937}.
The interaction can be decomposed into co-rotating terms $\propto(\hat{a}\hat{\sigma}_{+}+\hat{a}^{\dagger}\hat{\sigma}_{-})$,
which govern the standard Jaynes--Cummings (JC) interaction, and
counter-rotating terms $\propto(\hat{a}^{\dagger}\hat{\sigma}_{+}+\hat{a}\hat{\sigma}_{-})$,
which govern the anti--Jaynes--Cummings (AJC) interaction. While
the JC part conserves the standard excitation number $\hat{N}_{\text{JC}}=\hat{a}^{\dagger}\hat{a}+\frac{1}{2}(\hat{\sigma}_{z}+1)$,
the AJC interaction conserves $\hat{N}_{\text{AJC}}=\hat{a}^{\dagger}\hat{a}-\frac{1}{2}(\hat{\sigma}_{z}+1)$~\cite{BocanegraGaray2024}.
Although the individual counter-rotating processes change the conventional
excitation number by $\pm2$, both the JC and AJC interactions (and
the combined quantum Rabi Hamiltonian) strictly preserve the discrete
$\mathbb{Z}_{2}$ parity symmetry governed by $\hat{\Pi}=\exp[i\pi(\hat{a}^{\dagger}\hat{a}+\frac{1}{2}(\hat{\sigma}_{z}+1))]$~\cite{Braak2011,Baksic2014}.

The counter-rotating part can be omitted using the rotating--wave
approximation (RWA), when the cavity and qubit frequencies are near
resonance and the coupling is not too strong. The RWA simplifies the
Rabi model to the JC model which works well in the weak and strong
coupling regimes of cavity and circuit QED~\cite{Jaynes1963}. The
JC model has been the basic model to describe seminal demonstrations
such as quantum non--demolition qubit readout~\cite{Wallraff2004},
the generation of Fock states~\cite{Hofheinz2008}, the observation
of multiphoton resonances~\cite{Deppe2008}, and the cavity mediated
coupling between two qubits~\cite{Majer2007,Sillanpaeae2007}.

However, the RWA breaks down in the ultra-strong coupling regime where
the coupling strength $g$ becomes a significant fraction of the system
frequencies ($0.1\le g/\omega_{r}\le1$)~\cite{Niemczyk2010}, and
in the deep-strong coupling regime ($g/\omega_{r}\ge1$)~\cite{Casanova2010},
as well as in the strong driving regimes~\cite{TwyeffortIrish2007,Zueco2009}.
In these regimes, the counter-rotating terms must be retained in the
interaction Hamiltonian. The excitation-number nonconserving nature
of these terms leads to nontrivial phenomena absent in the JC model.
These include modified energy spectra observed as Bloch-Siegert shifts~\cite{Bloch1940,Tuorila2010,FornDiaz2010},
generation of Schr\"odinger cat states~\cite{Solano2003}, asymmetries
in vacuum Rabi splitting~\cite{Beaudoin2011}, multiphoton resonances~\cite{Ma2015},
modified non-RWA input-output relations and open-system response~\cite{Ciuti2006,Ridolfo2012,Shang2019,Mercurio2022,Shang2026},
quantum phase transitions~\cite{Hwang2015}, and the alteration
of photon blockade physics~\cite{Hwang2016,LeBoite2016,Shen2018,Trivedi2019}.
Beyond static effects, the interplay of JC and AJC terms in the generalized
Rabi dynamics significantly impacts coherent energy transfer~\cite{Lu2025},
the quantum Zeno effect~\cite{Zheng2008}, and entanglement evolution
in open systems~\cite{Ficek2002,Jing2009,Chen2010,Shang2024}, while
playing a crucial role in the dynamics of driven two-level systems~\cite{Lue2012,Grimsmo2013}.

These non-RWA effects have been extensively explored and experimentally
simulated across diverse architectures~\cite{Ashhab2007,FornDiaz2010,Ballester2012,Deng2015,Yoshihara2017,Braumueller2017,Langford2017},
and utilized for Floquet engineering and ultra-fast, high-fidelity
geometric quantum operations~\cite{Chen2022,Akbari2025}. Furthermore,
parametric modulation and on-demand coupling schemes have been developed
to transition between light-{}-matter interaction regimes~\cite{Valadares2024}
and to engineer both red-sideband (JC-like)~\cite{Beaudoin2012,Strand2013,Wang2019}
and blue-sideband (AJC-like) interactions~\cite{Wallraff2007,Leek2009,Novikov2015,Lu2017},
with the latter acting as a coherent two-photon pump. Emulation of
these models in systems like modulated qutrits~\cite{Dodonov2019}
and the demonstration of pure AJC effects in dipole-dipole coupling~\cite{Wang2017}
further highlight the broad relevance of these interactions. Recent
studies show that the AJC model with squeezed light exhibits distinct
photon statistics and enhanced entanglement~\cite{Mayero2024},
while switching between JC and AJC dynamics can control quantum revivals~\cite{Lara2005}.

A crucial observation is that these phenomena depend on the relative
weight of the JC and AJC interaction channels. In conventional circuit-QED
architectures with purely capacitive or inductive coupling, the relative
strengths of these two channels are fixed by the circuit geometry
and the qubit--resonator parameters. Tunable couplers have nevertheless
enabled highly controllable qubit--qubit interactions and high-fidelity
gates~\cite{Yan2018}, while the anisotropic Rabi model (ARM) provides
a theoretical framework in which the relative weights of the JC and
AJC terms can be varied~\cite{Xie2014}. The broader physics of
counter-rotating interactions and their control has also been extensively
explored, including through ultra-strong coupling architectures~\cite{FriskKockum2019,FornDiaz2019}
and symmetry engineering~\cite{Baksic2014}. More recently, the
tunable JC/AJC structure of the ARM has been considered for noise-biased
qubit designs~\cite{Yu2026}. These studies establish both the importance
of controlling the two interaction channels and the feasibility of
tunable superconducting couplers. However, a circuit architecture
that directly and in situ controls the relative JC and AJC contributions
through the same physical coupling element, without relying on strong
parametric driving or restricting the system to a fixed interaction
anisotropy, remains desirable. This distinction is important because
controlling the overall qubit--resonator coupling strength is not
equivalent to independently controlling the balance between the co-rotating
and counter-rotating channels.

In this work, we propose a circuit that achieves this goal. By integrating
a flux-tunable coupler that mediates the capacitive interaction between
the qubit and the resonator~\cite{Yan2018}, while the qubit couples
simultaneously and directly to the resonator’s voltage antinode (capacitively)
and current antinode (inductively), we demonstrate how the relative
amplitudes of the JC and AJC interaction channels can be controlled
in situ by the external flux through the coupler. In contrast to previous
tunable-coupling schemes, where tunability primarily refers to the
overall interaction strength, our architecture exploits the coexistence
of capacitive and inductive coupling paths to tune the anisotropy
of the light--matter interaction. This enables the system to be dynamically
tuned across the full parameter space of the ARM, from pure JC to
pure AJC regimes. Crucially, the pure JC interaction can be reached
without invoking the conventional RWA, while the pure AJC interaction
can be realized without requiring external parametric modulation.

\section*{Circuit Model}

\begin{figure}
\begin{centering}
\includegraphics[width=10cm]{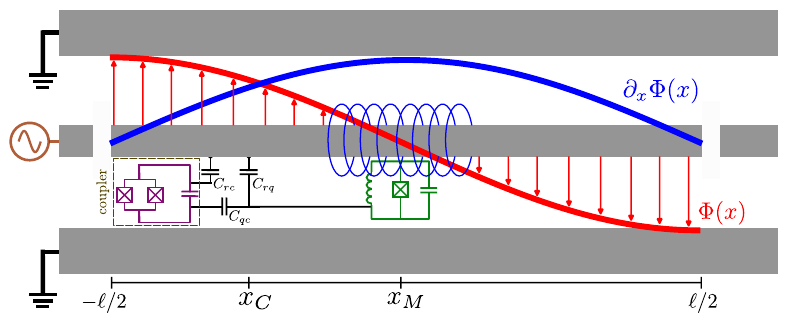}
\par\end{centering}
\caption{Schematic of the proposed circuit featuring a tunable coupling architecture.
A superconducting qubit (green) is inductively coupled to a resonator
at position $x_{M}$, at which the current profile of the resonator
 which is proportional to $\partial_{x}\Phi(x,t)$ (blue profile), shows
an antinode. The qubit is simultaneously connected to the resonator
both directly (via the cross-capacitance $C_{rq}$) and indirectly
through a tunable coupler (purple box) at position $x_{C}$, chosen
near an antinode of the resonator's voltage, which is proportional
to $\partial_{t}\Phi(x,t)$ (red profile). The coupler mediates the interaction
via capacitances $C_{rc}$ and $C_{qc}$. By tuning the external magnetic
flux threading the coupler, the effective qubit-resonator coupling
can be dynamically adjusted.}\label{fig:1}
\end{figure}

A schematic of the proposed circuit is shown in Fig.~\ref{fig:1}.
The system comprises three components: a microwave resonator ($r$),
a superconducting qubit ($q$), and an intermediate flux-tunable transmon
coupler ($c$). The qubit is coupled to the resonator via three distinct
pathways: a direct inductive coupling (with strength $g^{(L)}_{rq}$),
a direct capacitive coupling ($g^{(C)}_{rq}$), and an indirect capacitive
interaction mediated by a flux-tunable transmon coupler.

The total Hamiltonian of the circuit is given by (see Appendix for
detailed derivation) 
\begin{equation}
H=H_{0}+H_{\mathrm{int}},\label{eq:H_full-1}
\end{equation}
where the uncoupled Hamiltonian reads ($\hbar=1$)
\begin{equation}
H_{0}=\omega_{r}\hat{a}^{\dagger}\hat{a}+\frac{\omega_{q}}{2}\hat{\sigma}_{z}+\omega_{c}\hat{c}^{\dagger}\hat{c}-\frac{\alpha_{c}}{2}\hat{c}^{\dagger}\hat{c}^{\dagger}\hat{c}\hat{c},
\end{equation}
with $\hat{a}$ ($\hat{c}$) being the annihilation operator of the
resonator (coupler), $\hat{\sigma}_{z}$ the Pauli-$Z$ operator of
the qubit, and $\alpha_{c}$ the anharmonicity of the transmon coupler.
Operating in the low-energy regime where the coupler remains in its
two lowest energy states, the full interaction Hamiltonian takes the
form
\begin{align}
H_{\mathrm{int}}= & ig^{C}_{rq}(\hat{a}^{\dagger}-\hat{a})\hat{\sigma}_{y}-g^{L}_{rq}(\hat{a}^{\dagger}+\hat{a})\hat{\sigma}_{x}\nonumber \\
 & -g^{C}_{rc}(\hat{a}^{\dagger}-\hat{a})(\hat{c}^{\dagger}-\hat{c})+ig^{C}_{qc}(\hat{c}^{\dagger}-\hat{c})\hat{\sigma}_{y},
\end{align}
where $g^{C}_{ij}$ and $g^{L}_{rq}$ denote the bare capacitive and
inductive coupling strengths, respectively.

When the coupler is strongly detuned from both the resonator and the
qubit modes, i.e., $|\omega_{c}-\omega_{r}|,|\omega_{c}-\omega_{q}|\gg g^{C}_{rc},g^{C}_{qc}$,
the coupler degrees of freedom can be adiabatically eliminated using
a second-order Schrieffer-Wolff (SW) transformation (see Appendix
for details). Projecting the transformed Hamiltonian onto the coupler
ground state subspace yields the effective bipartite qubit-resonator
Hamiltonian
\begin{equation}
H_{\mathrm{eff}}=\tilde{\omega}_{r}\hat{a}^{\dagger}\hat{a}+\frac{\tilde{\omega}_{q}}{2}\hat{\sigma}_{z}+i\tilde{g}^{C}_{rq}(\hat{a}^{\dagger}-\hat{a})\hat{\sigma}_{y}-g^{L}_{rq}(\hat{a}^{\dagger}+\hat{a})\hat{\sigma}_{x},\label{eq:Heff}
\end{equation}
where the Lamb-shifted mode frequencies are
\begin{align}
\tilde{\omega}_{r} & =\omega_{r}+(g^{C}_{rc})^{2}\left(\frac{1}{\omega_{r}-\omega_{c}}-\frac{1}{\omega_{r}+\omega_{c}}\right),\\
\tilde{\omega}_{q} & =\omega_{q}+(g^{C}_{qc})^{2}\left(\frac{1}{\omega_{q}-\omega_{c}}+\frac{1}{\omega_{q}+\omega_{c}}\right),
\end{align}
and the total effective capacitive coupling strength $\tilde{g}^{C}_{rq}$
consists of the direct and coupler-mediated interaction terms
\begin{equation}
\tilde{g}^{C}_{rq}=g^{C}_{rq}+g^{C}_{\mathrm{ind}},
\end{equation}
with the mediated capacitive interaction given by
\begin{equation}
g^{C}_{\mathrm{ind}}=\frac{1}{2}g^{C}_{rc}g^{C}_{qc}\left(\frac{1}{\omega_{r}-\omega_{c}}-\frac{1}{\omega_{r}+\omega_{c}}+\frac{1}{\omega_{q}-\omega_{c}}-\frac{1}{\omega_{q}+\omega_{c}}\right).\label{eq:g_ind}
\end{equation}

\section*{In-Situ Tuning of Interaction Regimes}

\begin{figure}
\begin{centering}
\includegraphics[width=16cm]{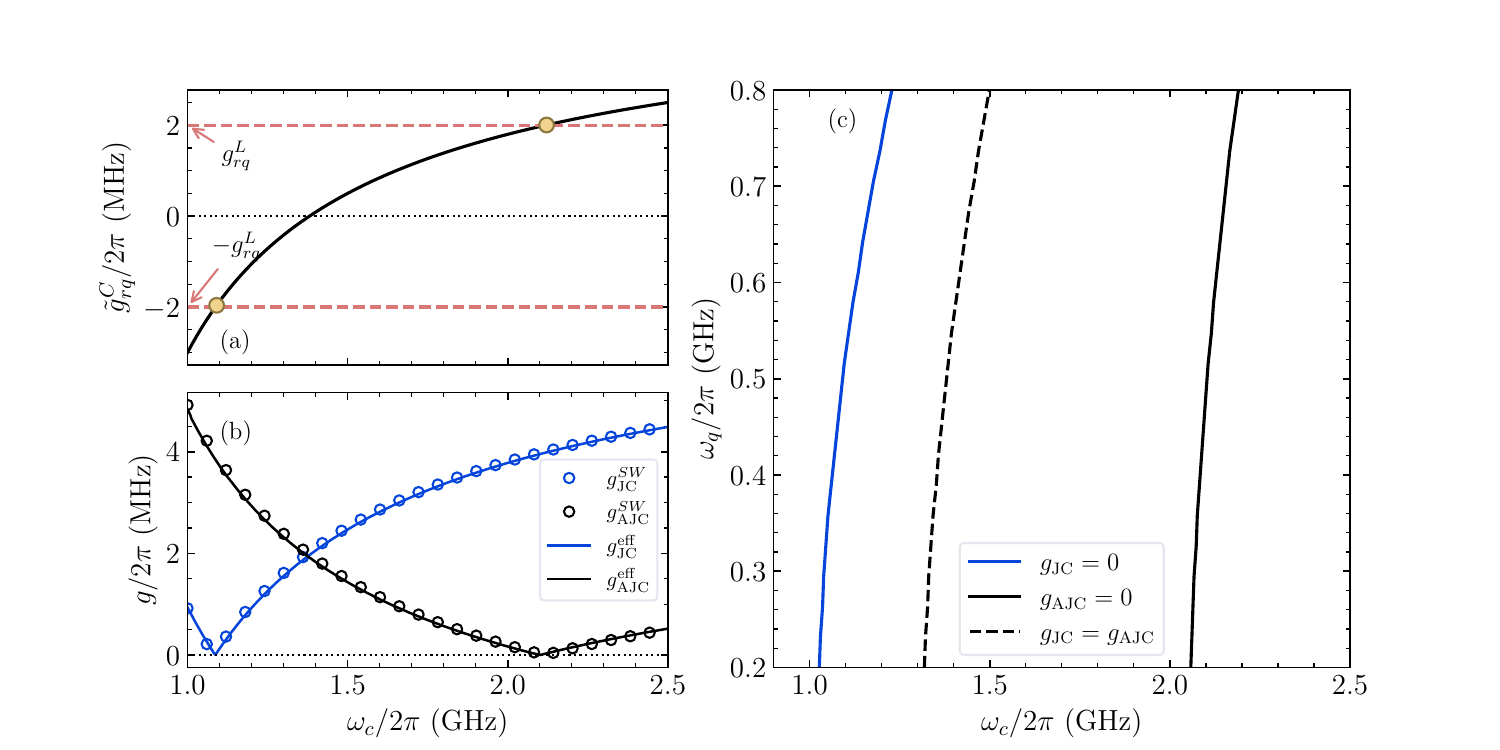}
\par\end{centering}
\caption{(a) Coupler-mediated capacitive coupling strength $\tilde{g}^{C}_{rq}$
as a function of the coupler frequency $\omega_{c}$. Horizontal dashed
lines indicate the critical cancellation conditions $\tilde{g}^{C}_{rq}=\pm g^{L}_{rq}$.
(b) Solid lines show analytical second-order SW transformation results
($g^{\mathrm{SW}}_{\mathrm{JC}},g^{\mathrm{SW}}_{\mathrm{AJC}}$)
given in Eqs.~\ref{eq:g_JC} and \ref{eq:g_AJC}. Circles show effective
coupling strengths ($g^{\mathrm{ED}}_{\mathrm{JC}},g^{\mathrm{ED}}_{\mathrm{AJC}}$)
obtained using numerical ED of the full tripartite Hamiltonian. (c)
The pure AJC ($g_{\mathrm{JC}}=0$), standard Rabi ($g_{\mathrm{JC}}=g_{\mathrm{AJC}}$),
and pure JC ($g_{\mathrm{AJC}}=0$) regimes are shown in $(\omega_{c},\omega_{q})$
plane. In (a) and (b) the qubit frequency is fixed $\omega_{q}/2\pi=0.5\,\mathrm{GHz}$.
Other parameters are $\omega_{r}/2\pi=0.5\,\mathrm{GHz}$, $\alpha_{c}/2\pi=0.2\,\mathrm{GHz}$,
$g^{L}_{rq}/2\pi=2\,\mathrm{MHz}$, $g^{C}_{rq}=5\,\mathrm{MHz}$,
and $g^{C}_{rc}/2\pi=20\,\mathrm{MHz},\,g^{C}_{qc}/2\pi=150\,\mathrm{MHz}$.}\label{fig:couplings_sweep}
\end{figure}

The underlying mechanism enabling the continuous tunability of the
effective qubit-resonator interaction is illustrated in Fig.~\ref{fig:couplings_sweep}(a),
which displays the dependence of the effective capacitive coupling
$\tilde{g}^{C}_{rq}$ on the bare coupler frequency $\omega_{c}$.
The qubit frequency is chosen $\omega_{q}/2\pi=0.5\,\mathrm{GHz}$
which is typical to fluxonium-type qubits~\cite{Nguyen2019}. Also,
the transmon tunable coupler operates at a higher frequency range
($\omega_{c}/2\pi=1\sim2.5\,\mathrm{GHz}$), to ensure its dispersive
interaction with the qubit and the resonator~\cite{Yan2018}.

In-situ tuning of $\omega_{c}$ via the external magnetic flux modifies
$g^{C}_{\mathrm{ind}}$, causing $\tilde{g}^{C}_{rq}$ to vary monotonically
and undergo a sign reversal across the parameter space. In particular,
at the coupler frequencies marked by solid circles in Fig.\ref{fig:couplings_sweep}(a),
the effective capacitive coupling $\tilde{g}^{C}_{rq}$ takes values
$\pm g^{L}_{rq}$, which as we discuss in the following, provides
the basis for the interference mechanism leading to the anisotopic
Rabi model in the system.

Although either of the capacitive ($\tilde{g}^{C}_{rq}$) or the inductive
coupling ($g^{L}_{rq}$) in Eq.(\ref{eq:Heff}) leads to an effective
JC interaction under the RWA, their simultaneous presence is more
interesting without invoking this approximation. In this case, the
interaction part of Eq.(\ref{eq:Heff}) can be rearranged into JC
and AJC interactions 
\begin{equation}
H^{\mathrm{eff}}_{\mathrm{int}}=g_{\mathrm{JC}}(\hat{a}^{\dagger}\hat{\sigma}_{-}+\hat{a}\hat{\sigma}_{+})+g_{\mathrm{AJC}}(\hat{a}^{\dagger}\hat{\sigma}_{+}+\hat{a}\hat{\sigma}_{-}),\label{eq:H_aniso_Rabi}
\end{equation}
where the effective JC and AJC coupling strengths are 
\begin{align}
g_{\mathrm{JC}} & =-\left(\tilde{g}^{C}_{rq}+g^{L}_{rq}\right)=-\left(g^{C}_{rq}+g^{C}_{\mathrm{ind}}(\omega_{c})+g^{L}_{rq}\right),\label{eq:g_JC}\\
g_{\mathrm{AJC}} & =\tilde{g}^{C}_{rq}-g^{L}_{rq}=g^{C}_{rq}+g^{C}_{\mathrm{ind}}(\omega_{c})-g^{L}_{rq}.\label{eq:g_AJC}
\end{align}
It is seen that by controlling the ratio $g^{L}_{rq}/\tilde{g}^{C}_{rq}$,
the two coupling channels can interfere, allowing one to enhance or
suppress either the JC or AJC interaction channels. This is precisely
enabled by the active tuning of the coupler frequency $\omega_{c}$,
which dynamically adjusts $\tilde{g}^{C}_{rq}$, as described above.
Note that the overall minus sign in Eq.(\ref{eq:g_JC}) is purely
a global phase convention originating from the definition of the capacitive
and inductive dipole operators, and carries no dynamical consequences.

Fig.~\ref{fig:couplings_sweep}(b) displays the resulting effective
JC and AJC coupling strengths ($g_{\mathrm{JC}}$ and $g_{\mathrm{AJC}}$)
as a function of $\omega_{c}$. We compare the analytical second-order
SW transformation predictions ($g^{\mathrm{SW}}_{\mathrm{JC}},g^{\mathrm{SW}}_{\mathrm{AJC}}$)
against the coupling strengths obtained from numerical exact diagonalization
(ED) of the full tripartite Hamiltonian ($g^{\mathrm{ED}}_{\mathrm{JC}},g^{\mathrm{ED}}_{\mathrm{AJC}}$)
(see Appendix for details). The analytical and numerical results are
in perfect agreement.

As shown in Fig.~\ref{fig:couplings_sweep}(b), at the specific coupler
frequency where $\tilde{g}^{C}_{rq}=-g^{L}_{rq}$, the co-rotating
component (and thus $g_{\mathrm{JC}}$) vanishes, yielding a pure
AJC interaction Hamiltonian $H_{\mathrm{int}}=g_{\mathrm{AJC}}(\hat{a}^{\dagger}\hat{\sigma}_{+}+\hat{a}\hat{\sigma}_{-})$
without requiring active parametric drives. On the other hand, when
$\tilde{g}^{C}_{rq}=g^{L}_{rq}$, counter-rotating processes vanish
($g_{\mathrm{AJC}}=0$), realizing the standard JC interaction $H_{\mathrm{int}}=g_{\mathrm{JC}}(\hat{a}^{\dagger}\hat{\sigma}_{-}+\hat{a}\hat{\sigma}_{+})$.
At the points in between, when the coupler frequency is tuned such
that the net capacitive coupling vanishes ($\tilde{g}^{C}_{rq}=0$),
the interaction becomes purely inductive, yielding the standard isotropic
Rabi interaction $H_{\mathrm{int}}=-g^{(L)}_{rq}(\hat{a}^{\dagger}+\hat{a})\hat{\sigma}_{x}$.

A phase diagram in the $(\omega_{c},\omega_{q})$ parameter space
is also shown in Fig.~\ref{fig:couplings_sweep}(c) to mark the regions
where pure AJC, pure JC and standard Rabi models are expected. Using
this diagram distinct coupling regimes can be selectively targeted
by adjusting $\omega_{c}$ and $\omega_{q}$.

\section*{Numerical Results}

In our numerical simulations, the Hilbert spaces of the resonator
and the coupler are truncated at Fock basis dimensions $N_{r}=8$
and $N_{c}=8$, respectively, which ensures numerical convergence
across the parameter regimes considered. Dissipative dynamics are
modeled via the Lindblad master equation with bare resonator decay
rate $\kappa/2\pi=0.1\,\text{MHz}$ and coupler decay rate $\gamma_{c}/2\pi=0.01\,\text{MHz}$.
For the qubit mode, we take a very small intrinsic relaxation rate
($\gamma_{q}/2\pi\sim0.1\,\text{kHz}$). The numerical calculations
are performed using \textsc{QuTiP} package~\cite{Lambert2026}.

\subsection*{Cavity Transmission Spectroscopy and Dispersive Shift}

To characterize the spectroscopic signatures of the anisotropic light--matter
interaction, we consider the linear-response spectrum of the resonator
under a weak probe. See appendix for details of calculating the spectrum.
Fig.~\ref{fig:spect} shows the resulting spectrum as a function
of the qubit frequency $\omega_{q}$ and the probe frequency $\omega_{p}$
for three representative coupler frequencies: $\omega_{c}/2\pi=1.09,\,1.4$
and $2.1\,{\rm GHz}$, which are corresponding to pure AJC, isotropic
Rabi, and pure JC limits, respectively.

\begin{figure}
\begin{centering}
\includegraphics[width=16cm]{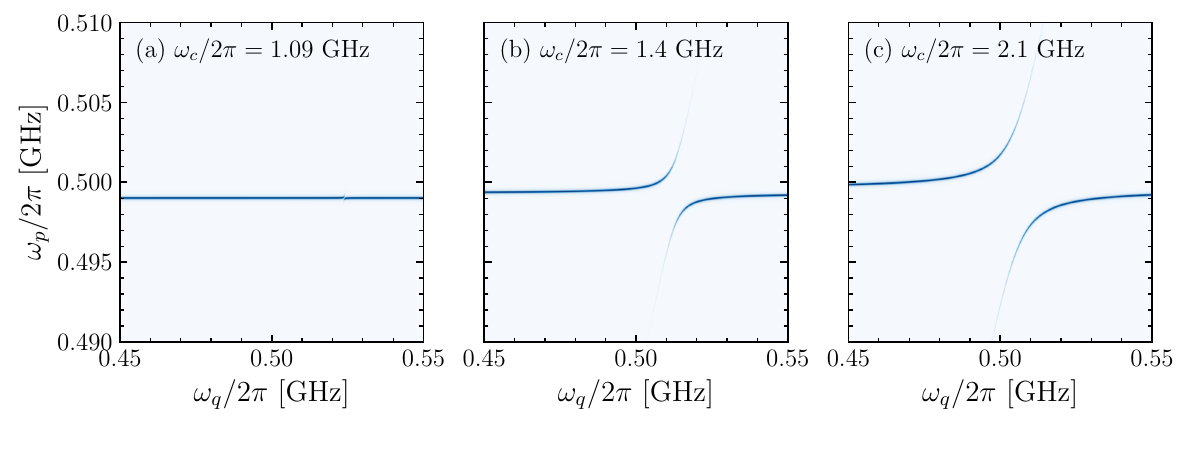}
\par\end{centering}
\caption{The resonator transmission as a function of probe frequency $\omega_{p}$
and qubit frequency $\omega_{q}$, for different $\omega_{c}$ values:
(a) $\omega_{c}/2\pi=1.09\,{\rm GHz}$ (AJC regime), (b) $\omega_{c}/2\pi=1.4\,{\rm GHz}$
(Rabi regime), and (c) $\omega_{c}/2\pi=2.1\,{\rm GHz}$ (JC regime).
All parameters are as in Fig.~\ref{fig:couplings_sweep}, and $\ensuremath{\kappa/2\pi=0.1\,\mathrm{MHz}}$.}\label{fig:spect}
\end{figure}

In the pure AJC regime {[}Fig.~\ref{fig:spect}(a){]}, the spectrum
exhibits a single dispersionless resonance at the renormalized resonator
frequency $\omega_{p}\simeq\tilde{\omega}_{r}$. In particular, no
appreciable vacuum Rabi splitting is observed as $\omega_{q}$ is
tuned through the resonator frequency. This behavior follows from
the structure of the AJC interaction: the counter-rotating coupling
predominantly connects states whose bare transition frequencies are
far detuned in the vicinity of $\omega_{q}\simeq\omega_{r}$. Consequently,
it does not produce the avoided crossing which is responsible for
the conventional vacuum Rabi splitting.

For the isotropic Rabi regime shown in Fig.~\ref{fig:spect}(b) and
the pure JC regime shown in Fig.~\ref{fig:spect}(c), the situation
changes qualitatively. As the qubit frequency approaches the resonator
frequency, the single resonance evolves into two distinct branches,
signaling the formation of hybridized light--matter eigenstates.
The separation of the two branches at resonance is approximately $2g_{{\rm JC}}$,
where $g_{{\rm JC}}$ denotes the JC coupling strength in Eq.(\ref{eq:H_aniso_Rabi}).
Thus, the spectrum directly reveals the coherent hybridization induced
by the resonant part of the light--matter interaction. Importantly,
the value of $g_{{\rm JC}}$ in these different panels is not a fixed
parameter and is determined by the coupler frequency, which continuously
tunes the relative strengths of the JC and AJC components of the effective
anisotropic Rabi interaction.

Taken together, these spectra provide a direct spectroscopic signature
of the tunability of the effective light--matter interaction. By
varying $\omega_{c}$, the relative weights of the JC and AJC processes
can be continuously modified, allowing the system to evolve from a
regime in which the conventional vacuum Rabi splitting is absent to
regimes exhibiting the usual avoided-crossing structure.

Now we focus on the dispersive regime in the effective qubit-resonator
subsystem, where the detuning $\left|\omega_{q}-\omega_{r}\right|$
significantly exceeds the effective couplings. By performing a second-order
SW transformation on the qubit-resonator subsystem, the state-dependent
dispersive shift $\chi$ is obtained as (see Apeendix for derivation)
\begin{equation}
\chi^{\mathrm{SW}}=\frac{g^{2}_{\mathrm{JC}}}{\Delta}+\frac{g^{2}_{\mathrm{AJC}}}{\Sigma},\label{eq:chi_dispersive}
\end{equation}
where $\Delta\equiv\omega_{q}-\omega_{r}$ is the qubit-resonator
detuning and $\Sigma\equiv\omega_{q}+\omega_{r}$. Here, $\chi$ parameterizes
the dispersive frequency shift per qubit excitation, corresponding
to a total cavity pull of $2\chi$ between the ground and excited
states. In the negative-detuning regime ($\Delta<0$), the JC and
AJC contributions carry opposite signs. Consequently, tuning the coupler
frequency $\omega_{c}$ allows one to continuously vary the relative
weights of $g_{\mathrm{JC}}$ and $g_{\mathrm{AJC}}$, driving the
net dispersive shift through zero ($\chi=0$) at a critical sweet-spot
frequency $\omega_{c}=\bar{\omega}_{c}$ satisfying $g_{\mathrm{AJC}}/g_{\mathrm{JC}}=\sqrt{\Sigma/|\Delta|}$.

\begin{figure}
\begin{centering}
\includegraphics[width=14cm]{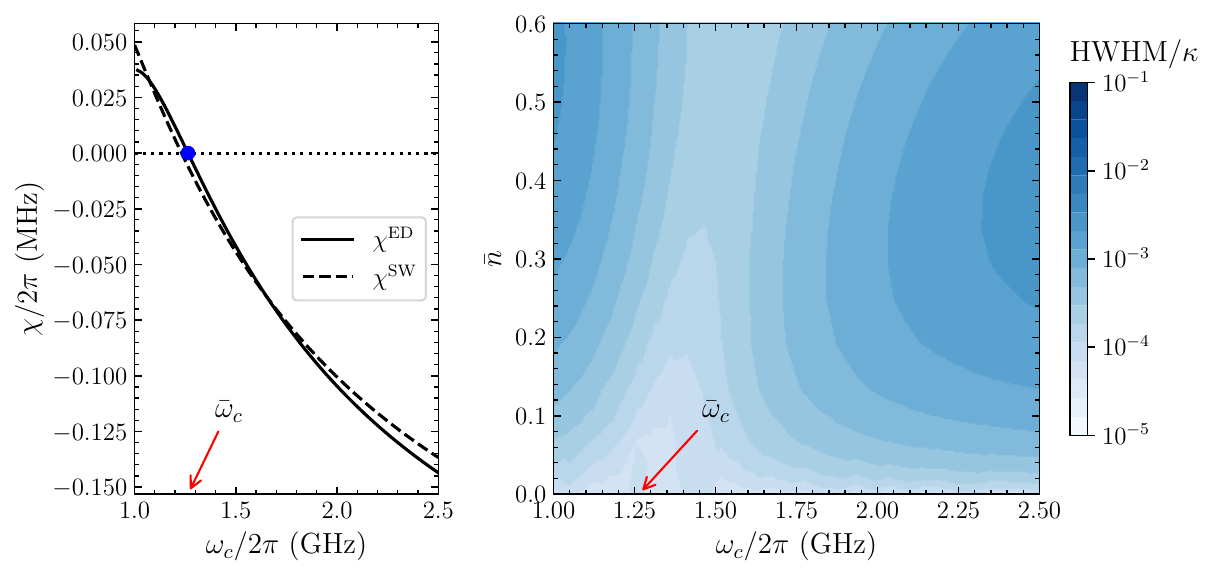}
\par\end{centering}
\caption{(a) Dispersive shift $\chi$ as a function of coupler frequency $\omega_{c}$,
comparing the numerical ED $\chi^{\mathrm{ED}}$ (solid line) with
the analytical perturbative result $\chi^{\mathrm{SW}}$ (dashed line).
The sweet spot $\bar{\omega}_{c}$ marks the exact cancellation point
$\chi=0$. (b) Normalized spectral linewidth $\mathrm{HWHM}/\kappa$
of the qubit as a function of $\omega_{c}$ and average photon number
$\bar{n}$, highlighting the photon-noise-insensitive operating regime
around $\bar{\omega}_{c}$. Here the qubit frequency is $\omega_{q}/2\pi=0.2\,\mathrm{GHz}$,
the resonator's decay rate is $\ensuremath{\kappa/2\pi=0.1\,\mathrm{MHz}}$,
and the coupler's decay rate is $\ensuremath{\gamma_{c}/2\pi=0.01\,\mathrm{MHz}}$
All other parameters are as in Fig.~\ref{fig:couplings_sweep}.}\label{fig:chi}
\end{figure}

To validate this analytical expression, we compare $\chi^{\mathrm{SW}}$
with the dispersive shift $\chi^{\mathrm{ED}}$ obtained via full
numerical diagonalization of the tripartite Hamiltonian, defined as
$2\chi^{\mathrm{ED}}=(E_{e,1,0}-E_{g,1,0})-(E_{e,0,0}-E_{g,0,0})$,
where $E_{m,n,o}$ is the energy of the eigenstate in which the qubit
is in state $m$ and $(n,o)$ stand for the number of excitations
in the resonator and coupler, respectively. As shown in Fig.~\ref{fig:chi}(a),
$\chi^{\mathrm{SW}}$ matches $\chi^{\mathrm{ED}}$ with good accuracy
across the entire frequency range. As $\omega_{c}$ increases, $\chi$
decreases monotonically from positive values in the AJC-dominated
regime, crosses zero at $\omega_{c}=\bar{\omega}_{c}$, and becomes
increasingly negative as the JC interaction becomes dominant.

\subsection*{Photon-Shot-Noise Dephasing and Purcell Decay Suppression}

The vanishing of $\chi$ at $\bar{\omega}_{c}$ has profound implications
for mitigating qubit dephasing. In an idealized bipartite qubit-resonator
system within the dispersive regime, cavity photon-number fluctuations
induce photon shot-noise dephasing at a rate given by~\cite{Schuster2005,Blais2021}
\begin{equation}
\Gamma_{\phi}=\frac{2\kappa\chi^{2}\bar{n}}{\Delta^{2}+\chi^{2}+(\kappa/2)^{2}},\label{eq:gamma_phi}
\end{equation}
where $\kappa$ is the resonator decay rate and $\bar{n}$ is the
average intra-cavity thermal photon population. According to Eq.~(\ref{eq:gamma_phi}),
setting $\chi=0$ completely suppresses this leading-order cavity
shot-noise channel.

However, Eq.~(\ref{eq:gamma_phi}) strictly applies to a purely bipartite
system and neglects the dissipation and dephasing channels intrinsic
to our tripartite (qubit-coupler-resonator) architecture. In the tripartite
system, the presence of the tunable coupler opens additional dephasing
and decoherence pathways. First, low-frequency magnetic flux fluctuations
($\delta\Phi_{\mathrm{ext}}$) threading the SQUID loop of the transmon
coupler cause direct fluctuations in the coupler frequency $\omega_{c}$.
Because the dressed qubit and resonator frequencies ($\tilde{\omega}_{q},\tilde{\omega}_{r}$)
are dispersively shifted through level repulsion with the coupler,
these variations translate into dressed qubit frequency fluctuations
via $\delta\tilde{\omega}_{q}\approx(\partial\tilde{\omega}_{q}/\partial\omega_{c})\delta\omega_{c}$,
thereby inducing an additional pure dephasing channel. Second, the
finite energy decay rate of the coupler, $\gamma$, induces Purcell-like
relaxation and dephasing on the dressed qubit through residual state
hybridization. Lastly, virtual transitions involving higher excited
states of the transmon coupler and multi-photon exchanges lead to
non-perturbative corrections that fall outside the second-order bipartite
model.

To comprehensively capture these multi-channel dephasing and dissipation
processes beyond both the bipartite and dispersive approximations,
we analyze the full tripartite open quantum system described by the
master equation. Using the quantum regression theorem, we compute
the steady-state qubit emission spectrum from the Fourier transform
of the two-time correlation function
\begin{equation}
S(\omega)=\mathrm{Re}\int^{\infty}_{0}\langle\sigma_{x}(t)\sigma_{x}(0)\rangle_{\mathrm{ss}}e^{-i\omega t}\,dt.\label{eq:spec_corr}
\end{equation}

To isolate the environmental dephasing and decoherence mediated exclusively
by the surrounding modes, we incorporate both cavity decay $\kappa$
and coupler dissipation $\gamma$, while setting the intrinsic bare
qubit relaxation and bare pure dephasing to zero. In this framework,
the half-width at half-maximum (HWHM) of the qubit spectral peak directly
quantifies the total transverse dephasing rate $\Gamma_{2}$ induced
collectively by cavity photon shot noise, coupler dissipation, and
hybrid-mode fluctuations~\cite{Schad2016}.

Figure~\ref{fig:chi}(b) displays the normalized HWHM as a function
of the coupler frequency $\omega_{c}$ and the average cavity photon
number $\bar{n}$. In the vicinity of the sweet spot $\bar{\omega}_{c}$,
the spectral linewidth attains a pronounced minimum and remains virtually
insensitive to the cavity photon occupation $\bar{n}$. Away from
$\bar{\omega}_{c}$, the finite dispersive shift $\chi$ activates
cavity shot-noise dephasing, leading to a substantial broadening of
the qubit linewidth with increasing $\bar{n}$ and $\omega_{c}$.
The small residual linewidth observed at $\bar{\omega}_{c}$ is due
to the coupling to the dissipative coupler and higher-order dispersive
dressings, confirming that operating at the dispersive sweet spot
provides robust, multi-channel suppression of the cavity-induced dephasing.

We emphasize that the in-situ tunability in this model system resolves
the inherent trade-off between coherence protection and readout fidelity.
During qubit operations or idling periods, the system can be dynamically
tuned to the sweet spot ($\chi\approx0$) to eliminate leading-order
photon shot-noise dephasing. For state measurement, a fast flux pulse
can temporarily tune the system away from this sweet spot to restore
the dispersive shift ($\chi\neq0$), enabling high-fidelity readout
before returning to the protected regime.

Finally, we address the Purcell effect, which describes the qubit
relaxation induced by photon leakage from the resonator. To fundamentally
understand the underlying mechanism of the Purcell dip and its dependence
on the coupler parameters, it is instructive to decompose the total
effective Purcell decay rate into its JC and AJC contributions~\cite{Sete2014}
\begin{equation}
\Gamma_{\mathrm{P}}=\Gamma^{\mathrm{JC}}_{\mathrm{P}}+\Gamma^{\mathrm{AJC}}_{\mathrm{P}},\label{eq:purcell}
\end{equation}
where
\begin{align}
\Gamma^{\mathrm{JC}}_{\mathrm{P}} & =\frac{\kappa g^{2}_{\mathrm{JC}}}{\Delta^{2}+\kappa^{2}+g^{2}_{\mathrm{JC}}},\label{eq:purcell_jc}\\
\Gamma^{\mathrm{AJC}}_{\mathrm{P}} & =\frac{\kappa g^{2}_{\mathrm{AJC}}}{\Sigma^{2}+\kappa^{2}+2g^{2}_{\mathrm{AJC}}}.\label{eq:purcell_ajc}
\end{align}
In the conventional JC case, the relaxation rate is amplified by the
relatively small detuning $\Delta$ in the denominator. Conversely,
the AJC rate is heavily suppressed by the extremely large frequency
sum $\Sigma$.

\begin{figure}
\begin{centering}
\includegraphics[width=10cm]{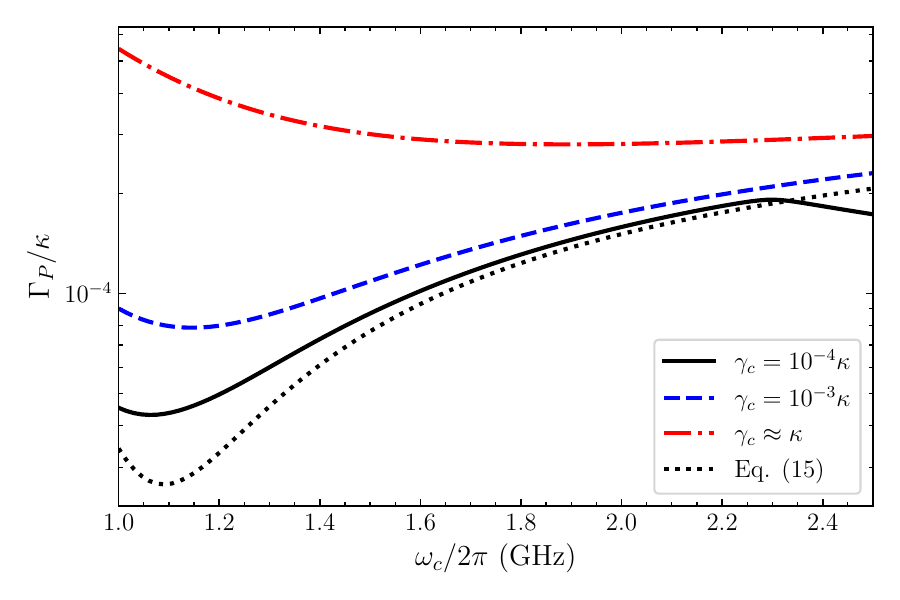}
\par\end{centering}
\caption{Normalized Purcell decay rate $\Gamma_{P}/\kappa$ of the qubit as
a function of the coupler frequency $\omega_{c}$, for different values
of coupler dissipation rates $\gamma_{c}$. The qubit frequency is
fixed at $\omega_{q}/2\pi=0.2$~GHz. The solid black, dashed blue,
and dash-dotted red curves correspond to $\gamma_{c}=10^{-4}\kappa$,
$10^{-3}\kappa$, and $\kappa$, respectively. Dotted curve is obtained
from Eq.~(\ref{eq:purcell}). All parameters are as in Fig.~\ref{fig:couplings_sweep}.}\label{fig:purcell}
\end{figure}

To rigorously calculate the Purcell decay rate ($\Gamma_{\mathrm{P}}$)
in the system, we diagonalize the full Liouvillian superoperator---which
incorporates the full tripartite Hamiltonian alongside the dissipation
channels of the resonator ($\kappa$) and the coupler ($\gamma_{c}$)---and
extract the relaxation rate from the real part of the first non-zero
eigenvalue. Since no intrinsic dissipation channel is assumed for
the bare qubit, the observed decay arises entirely from the leakage
of the qubit excitation through the lossy resonator and coupler modes.
This environment-induced leakage corresponds precisely to the Purcell
effect. The results obtained through this method are shown in Fig.~\eqref{fig:purcell},
where the Purcell decay rate $\Gamma_{P}$ of the qubit as a function
of the coupler frequency $\omega_{c}$, is plotted for different values
of coupler dissipation rates $\gamma_{c}$.

As illustrated previously in Fig.~\eqref{fig:couplings_sweep}(b),
the effective JC coupling $g_{\mathrm{JC}}$ vanishes around $\omega_{c}/2\pi\approx1.1$~GHz,
while the effective AJC coupling $g_{\mathrm{AJC}}$ remains prominent
at this frequency. Consequently, around this point, since the JC decay
channel is very small and the surviving pure AJC decay is heavily
suppressed by $\Sigma^{2}$, this regime yields a substantially reduced
overall Purcell decay rate. This mechanism manifests as the Purcell
dip observed for the nearly lossless coupler ($\gamma_{c}=10^{-4}\kappa$)
in Fig.~\eqref{fig:purcell}. For reference, we have plotted the
Purcell decay obtained for the effective qubit-resonator system given
by Eq. (\eqref{eq:purcell}) by the dotted curve in Fig.~\eqref{fig:purcell},
which can be considered as the Purcell decay when the coupler is dissipationless.

When $\gamma_{c}$ is increased in Fig.~\eqref{fig:purcell}, the
coupler opens a parallel, non-radiative leakage channel for the qubit.
The condition $g_{\mathrm{JC}}=0$ is no longer sufficient to fully
decouple the qubit from the environment, as relaxation can now occur
via the lossy coupler. To minimize the total decay rate in the presence
of coupler loss, the optimal operating point must shift toward larger
qubit-coupler detunings (higher $\omega_{c}$). Ultimately, in the
highly dissipative regime ($\gamma_{c}=\kappa$), the rapid photon
leakage from the coupler destroys the coherent interference required
to maintain $g_{\mathrm{JC}}=0$, washing out the pure AJC regime
entirely and resulting in a monotonically decreasing Purcell rate
with respect to $\omega_{c}$.

\section*{Noise Considerations}

The introduction of a flux-tunable transmon coupler provides \textit{in
situ} control over the effective capacitive coupling $\tilde{g}^{C}_{rq}$;
however, it inevitably exposes the system to external magnetic flux
noise $\delta\Phi_{\text{ext}}$ with a standard $1/f$ power spectrum
$S_{\Phi}(\omega)=A_{\Phi}/|\omega|$ ($A^{1/2}_{\Phi}\approx1-2\,\mu\Phi_{0}$~\cite{Ithier2005,Krantz2019}).
This flux noise induces fluctuations in the coupler transition frequency
$\omega_{c}(\Phi_{\text{ext}})$, which subsequently propagate into
the induced capacitive coupling $g^{C}_{\text{ind}}$ (Eq.~(\ref{eq:g_ind}))
and the dispersive shift $\chi$~(Eq.~(\ref{eq:chi_dispersive})).
It is thus important to derive the order of coupling fluctuations
and evaluate their impact on the core operational regimes.

For a symmetric SQUID loop with Josephson energy $E_{J,c}(\Phi_{\text{ext}})=E^{\max}_{J,c}|\cos(\pi\Phi_{\text{ext}}/\Phi_{0})|$,
the coupler frequency is $\omega_{c}\approx\sqrt{8E_{J,c}(\Phi_{\text{ext}})E_{C,c}}-E_{C,c}$.
A quasi-static flux fluctuation $\delta\Phi$ translates into a frequency
shift
\begin{equation}
\delta\omega_{c}=\frac{\partial\omega_{c}}{\partial\Phi_{\text{ext}}}\delta\Phi=-\frac{\pi}{2\Phi_{0}}\frac{\sqrt{8E^{\max}_{J,c}E_{C,c}}\sin(\pi\Phi_{\text{ext}}/\Phi_{0})}{\sqrt{|\cos(\pi\Phi_{\text{ext}}/\Phi_{0})|}}\,\delta\Phi.\label{eq:domega_c}
\end{equation}
Because the charge zero-point fluctuations scale as $n_{\text{zpf},c}\propto(E_{Jc}/E_{C,c})^{1/4}\propto\sqrt{\omega_{c}}$,
the direct interaction product scales linearly with frequency, $g^{C}_{rc}g^{C}_{qc}=\lambda\omega_{c}$,
where $\lambda$ is a geometric constant. Consequently, the fluctuation
in the effective capacitive coupling $\delta\tilde{g}^{C}_{rq}=\delta g^{C}_{\text{ind}}$
is given by $\delta\tilde{g}^{C}_{rq}=(\partial g^{C}_{\text{ind}}/\partial\omega_{c})\delta\omega_{c}$,
with 
\begin{equation}
\frac{\partial g^{C}_{\text{ind}}}{\partial\omega_{c}}=\frac{g^{C}_{\text{ind}}}{\omega_{c}}+\frac{1}{2}g^{C}_{rc}g^{C}_{qc}\sum_{j\in\{r,q\}}\left[\frac{1}{(\omega_{j}-\omega_{c})^{2}}-\frac{1}{(\omega_{j}+\omega_{c})^{2}}\right].\label{eq:dg_ind_domega}
\end{equation}
We observe that at the flux-insensitive sweet spot ($\Phi_{\text{ext}}=0$),
$\partial\omega_{c}/\partial\Phi_{\text{ext}}=0$, rendering the first-order
coupling fluctuations strictly zero ($\delta\tilde{g}^{C}_{rq}=0$).
On the other hand, at intermediate operating points away from the
sweet spot (e.g., $\Phi_{\text{ext}}\approx0.25-0.35\,\Phi_{0}$),
where the flux sensitivity reaches its maximum $|\partial\omega_{c}/\partial\Phi_{\text{ext}}|/2\pi\approx3-5\,\text{GHz}/\Phi_{0}$,
a standard flux noise amplitude of $\delta\Phi_{\text{rms}}=1\,\mu\Phi_{0}$
yields $\delta\omega_{c,\text{rms}}/2\pi\approx3-5\,\text{kHz}$.
Evaluating Eq.~\eqref{eq:dg_ind_domega} with our design parameters
($g^{C}_{rc}/2\pi=20\text{ MHz},\,g^{C}_{qc}/2\pi=150\,\text{MHz}$,
$\omega_{r}/2\pi=0.5\,\text{GHz}$, $\omega_{c}/2\pi\in[1.,2.5]\,\text{GHz}$)
gives $|\partial g^{C}_{\text{ind}}/\partial\omega_{c}|\approx(0.1-1.0)\times10^{-3}$,
which translates into an effective coupling fluctuation of 
\begin{equation}
\delta\tilde{g}^{C}_{rq}/2\pi\le10\,\text{kHz}.
\end{equation}

To achieve the pure JC or AJC regimes, we tune $\omega_{c}$ such
that $\tilde{g}^{C}_{rq}=g^{L}_{rq}$ or $\tilde{g}^{C}_{rq}=-g^{L}_{rq}$,
respectively. Compared to the nominal coupling strengths ($|g_{\text{JC}}|,|g_{\text{AJC}}|\sim5-10\,\text{MHz}$),
the residual coupling fluctuation $\delta\tilde{g}^{C}_{rq}$ constitutes
a relative deviation of less than $0.1\%$, confirming the structural
robustness of the pure JC and AJC operational points against $1/f$
flux noise.

It is also important to determine the order of fluctuations in the
dispersive shift $\chi$ induced by the magnetic flux fluctuations.
Around the cancellation sweet spot where $\chi\approx\frac{g^{2}_{{\rm JC}}}{\Delta}+\frac{g^{2}_{{\rm AJC}}}{\Sigma}=0$,
fluctuations in the effective capacitive coupling lead to
\begin{equation}
\delta\chi=\frac{\partial\chi}{\partial\tilde{g}^{C}_{rq}}\delta\tilde{g}^{C}_{rq}=2\left(\frac{g_{\text{JC}}}{\Delta}+\frac{g_{\text{AJC}}}{\Sigma}\right)\delta\tilde{g}^{C}_{rq}.
\end{equation}
Substituting our operational circuit parameters ($\Delta/2\pi\approx-300\,\text{MHz}$,
$\Sigma/2\pi\approx700\,\text{MHz}$, $g_{\text{JC}}/2\pi\approx1\,\text{MHz}$,
$g_{\text{AJC}}/2\pi\approx4\,\text{MHz}$) gives $|\partial\chi/\partial\tilde{g}^{C}_{rq}|\approx0.005$,
leading to a residual dispersive shift deviation of only $\delta\chi/2\pi\approx0.005\times10\,\text{kHz}\approx50\,\text{Hz}$.

We further consider the photon-shot-noise pure dephasing rate which
scales quadratically with the dispersive shift ($\Gamma_{\phi}\propto\chi^{2}\bar{n}$).
Thus, the residual dephasing is a second-order effect dictated by
the variance $\langle(\delta\chi)^{2}\rangle\propto\langle\delta\Phi^{2}\rangle$.
Consequently, even in the presence of flux noise outside $\Phi_{\text{ext}}=0$,
the residual dephasing rate remains orders of magnitude lower than
that of unmitigated dispersive architectures. We obtain
\begin{align}
\Gamma_{\phi,\text{residual}} & \propto\langle(\delta\chi)^{2}\rangle\bar{n}=\left(\frac{\partial\chi}{\partial\tilde{g}^{C}_{rq}}\frac{\partial g^{C}_{\text{ind}}}{\partial\omega_{c}}\frac{\partial\omega_{c}}{\partial\Phi_{\text{ext}}}\right)^{2}\langle\delta\Phi^{2}\rangle\bar{n}\nonumber \\
 & \approx2\pi\times(1.7\times10^{-5}\,\text{Hz})\bar{n},
\end{align}
which remains well below $1\,\text{mHz}$ even at $\bar{n}\le10$
intra-cavity photons, which is completely negligible compared to intrinsic
material dephasing ($\sim10-100\,\text{kHz}$). Moreover, since the
flux-noise frequency spread $\delta\omega_{c,\text{rms}}\approx5\,\text{kHz}$
is several orders of magnitude smaller than the characteristic width
of the HWHM dip ($\sim10-50\,\text{MHz}$ in Fig.~\eqref{fig:chi}),
the sharp coherence sweet spot is fully preserved without any perceptible
noise-induced smearing.

\section*{Practical Feasibility}

The simultaneous capacitive and inductive coupling architecture described
in this work can be implemented using standard superconducting circuit
technology. From a microscopic perspective, achieving this dual coupling
requires that both transition matrix elements $\langle g|\hat{n}|e\rangle$
and $\langle g|\hat{\varphi}|e\rangle$ be sufficiently nonzero. These
quantities measure the charge and flux participation of the qubit
transition, respectively. Under suitable external biases, both are
naturally finite in fluxonium qubits, ensuring that both coupling
mechanisms can coexist within realistic circuit parameters~(see Table~\ref{tab:circuit_parameters-1}).

To successfully sweep through pure JC and AJC interactions, our architecture
establishes a specific design constraint: $g^{C}_{rq}>g^{L}_{rq}$.
Because the tunable coupler frequency is dynamically operated above
the resonator and qubit frequencies ($\omega_{c}>\omega_{r},\omega_{q}$),
the induced coupling term $g^{C}_{\text{ind}}$ is strictly negative.
Therefore, by engineering a positive direct capacitive coupling $g^{C}_{rq}$
that exceeds the static inductive coupling $g^{L}_{rq}$, we ensure
that as $\omega_{c}$ is lowered towards $\omega_{r}$ (making $g^{C}_{\text{ind}}$
increasingly negative), the total effective coupling $\tilde{g}^{C}_{rq}$
sweeps downward, seamlessly intersecting both the $+g^{L}_{rq}$ and
$-g^{L}_{rq}$ thresholds.

From a fabrication standpoint, while both forms of coupling have been
extensively demonstrated individually in circuit QED, their integration
alongside a flux-tunable mediated pathway requires only modest extensions
of current practices~\cite{Sank2025,Chapple2025,Lin2025}. To establish
experimental verifiability, Table~\ref{tab:circuit_parameters-1}
provides a concrete lumped-element design case. The direct capacitive
couplings $C_{rq}\approx0.02\,\mathrm{fF}$, $C_{qc}\approx2.3\,\mathrm{fF}$
and $C_{rc}=35\,\mathrm{fF}$ are readily implemented via interdigital
capacitors, typically by positioning a coupling pad of the qubit in
close proximity to the voltage antinode of the resonator. The direct
mutual inductance $M_{rq}\approx35\,\mathrm{pH}$ is engineered through
a sub-micron shared geometric inductive trace, an approach successfully
demonstrated in strongly coupled flux~\cite{Wallraff2005,Yoshihara2017}
and fluxonium circuits~\cite{Manucharyan2009,Nguyen2019}.

\begin{table}[tb]
\begin{centering}
\begin{tabular}{lcc}
Parameter & Symbol & Physical Value\tabularnewline
\hline 
Resonator frequency & $\omega_{r}/2\pi$ & $0.5\text{ GHz}$\tabularnewline
Resonator characteristic impedance & $Z_{r}$ & $50\ \Omega$ ($\ensuremath{C_{r}\approx6.37\text{ pF},L_{r}\approx15.9\text{ nH}}$)\tabularnewline
Qubit charging energy & $E_{C}/h$ & $1.00\text{ GHz}$ ($C_{q}\approx19.3\text{ fF}$)\tabularnewline
Qubit Josephson energy & $E_{J}/h$ & $4.00\text{ GHz}$ ($L_{J}\approx40.7\text{ nH}$)\tabularnewline
Qubit inductive energy & $E_{L}/h$ & $0.50\text{ GHz}$ ($L_{\mathrm{arr}}\approx325.7\text{ nH}$)\tabularnewline
Qubit frequency & $\omega_{q}/2\pi$ & $0.23\text{ GHz}$\tabularnewline
Qubit number transition matrix element & $\langle g|\hat{n}|e\rangle$ & $0.1\text{--}0.5$\tabularnewline
Qubit flux transition matrix element & $\langle g|\hat{\varphi}|e\rangle$ & $0.4\text{--}2.5$\tabularnewline
Coupler max Josephson energy & $E^{\max}_{J,c}/h$ & $25.50\text{ GHz}$\tabularnewline
Coupler charging energy & $E_{C,c}/h$ & $0.13\text{ GHz}$ ($C_{c}\approx150\text{ fF}$)\tabularnewline
Coupler frequency tuning range & $\omega_{c}(\Phi_{c})/2\pi$ & $1.50\text{--}5.00\text{ GHz}$\tabularnewline
Direct coupling capacitance & $C_{rq}$ & $0.02\text{ fF}$ ($g^{C}_{rq}/2\pi\approx5\text{ MHz}$)\tabularnewline
Direct mutual inductance & $M_{rq}$ & $35\text{ pH}$ ($g^{L}_{rq}/2\pi\approx2\text{ MHz}$)\tabularnewline
Resonator-coupler capacitance & $C_{rc}$ & $35\text{ fF}$ ($g^{C}_{rc}/2\pi\approx20\text{ MHz}$)\tabularnewline
Qubit-coupler capacitance & $C_{qc}$ & $2.3\text{ fF}$ ($g^{C}_{qc}/2\pi\approx150\text{ MHz}$)\tabularnewline
\end{tabular}
\par\end{centering}
\caption{Circuit parameters for standard superconducting thin-film implementation
realizing the simultaneous capacitive-inductive coupling architecture.
Transition matrix elements and eigenfrequencies are calculated using
\textsc{scQubits} package~\cite{Groszkowski2021}.}\label{tab:circuit_parameters-1}
\end{table}

The qubit itself is chosen as a fluxonium, realized via a small Josephson
junction shunted by a high-kinetic-inductance array ($L_{\mathrm{arr}}\approx326\,\mathrm{nH},E_{L}/h=0.5\,\mathrm{GHz}$)~\cite{Hazard2019},
and operated near the half-flux sweet spot $\Phi_{q}=0.5\,\Phi_{0}$
to yield a transition frequency of $\omega_{q}/2\pi\approx0.2\text{--}0.5\,\mathrm{GHz}$.
The tunable coupler is formed by a standard SQUID-based transmon with
$E^{\max}_{J,c}/h=18.5\,\mathrm{GHz}$ and charging energy $E_{C,c}/h=130\,\mathrm{MHz}$.
Fabricating this tunable transmon involves standard lithography techniques
already perfected for multi-qubit processors~\cite{Yan2018,Marxer2023}.
These explicit circuit parameters rigorously yield the effective Hamiltonian
interaction strengths utilized throughout this work ($g^{C}_{rq}/2\pi=5\,\mathrm{MHz}$,
$g^{L}_{rq}/2\pi=3\,\mathrm{MHz}$, $g^{C}_{rc}/2\pi=20\,\mathrm{MHz}\text{ and }g^{C}_{qc}/2\pi=150\,\mathrm{MHz}$).

To clarify the operational domain, the present architecture operates
in the standard dispersive circuit QED regime with $g/\omega_{r}\sim10^{-2}$,
well below the ultra-strong coupling threshold. Within this domain,
the validity of the $\chi=0$ sweet spot beyond low-order perturbation
theory is rigorously confirmed by our non-perturbative numerical calculations.
Crucially, these numerical treatments account for the complete multi-level
transmon Hilbert space, confirming that the dispersive sweet spot
remains stable and is not an artifact of second-order SW approximations.

Regarding operational dynamics and robustness, the tuning speed of
the coupler is primarily limited by the flux line's bandwidth but
can typically be achieved on nanosecond timescales, which is orders
of magnitude faster than typical qubit coherence times~\cite{Chen2016,Sung2021}.
Furthermore, we acknowledge the effect of circuit asymmetries and
fabrication tolerances. Small mismatches between the intended capacitive
and inductive couplings will generically lead to an anisotropic Rabi
Hamiltonian with unbalanced JC and AJC contributions. However, this
does not eliminate the unique features of the proposed architecture.
Crucially, the in-situ adjustability of the mediated coupling $\tilde{g}^{C}_{rq}$
acts as a built-in continuous knob to actively fine-tune the system
post-fabrication. This ensures that pure interaction regimes (JC or
AJC) and sweet spots can be reliably reached and stabilized despite
typical fabrication variations, providing significant experimental
flexibility.

\section*{Conclusion}

In summary, we have introduced a circuit QED architecture capable
of implementing the full anisotropic Rabi model with dynamic, in-situ
tunability of its interaction symmetry. By combining a direct inductive
interaction with a flux-tunable capacitive channel mediated by a transmon
coupler, the constructive and destructive interference between the
two coupling pathways enables full control over the ratio $g_{{\rm AJC}}/g_{{\rm JC}}$.
This allows the system to seamlessly transition from the conventional
JC regime, through the isotropic Rabi model, to the pure AJC regime
without requiring parametric microwave drives or rotating-wave approximations.

We have analytically derived the effective Hamiltonian parameters
and validated them against numerical ED and generalized L\"owdin
downfolding. In the dispersive regime, we identified a distinct operational
sweet spot where the dispersive shift and the associated photon-shot-noise
dephasing can be suppressed. Crucially, in the pure AJC regime, the
Purcell decay rate is drastically suppressed while simultaneously
maintaining a high-contrast dispersive readout. Our noise sensitivity
and parameter tolerance analyses confirm that this design is fully
compatible with contemporary superconducting circuit fabrication processes.

Looking forward, this platform opens exciting prospects for both fundamental
quantum optics and quantum information processing. The ability to
dynamically engineer counter-rotating interactions in a stationary
circuit provides a versatile testbed for simulating non-equilibrium
quantum phase transitions, exploring novel parity-symmetry-breaking
phenomena, and generating non-classical photon states such as Schr\"odinger
cat states. Furthermore, extending this interference-based coupling
scheme to multi-qubit architectures holds strong promise for realizing
protected superconducting qubits and fast, tunable multi-qubit entangling
gates with intrinsic error suppression.

\section*{Appendix}

\section*{Derivation of circuit Hamilton}

\subsection*{Total Lagrangian}

We start from deriving the Lagrangian of the circuit diagram shown
in Fig.~\ref{fig:1}. The total Lagrangian of the circuit can be
expressed as the sum of the Lagrangians of its individual components
and their interactions: 
\begin{equation}
\mathcal{L}=\mathcal{L}_{{\rm res}}+\mathcal{L}_{{\rm qubit}}+\mathcal{L}_{{\rm coupler}}+\mathcal{L}_{{\rm {\rm C}}}+\mathcal{L}_{{\rm {\rm M}}}
\end{equation}
where $\mathcal{L}_{{\rm res}}$ describes the resonator, $\mathcal{L}_{{\rm qubit}}$
describes the qubit, $\mathcal{L}_{{\rm coupler}}$ describes the
tunable coupler, and $\mathcal{L}_{{\rm C}}$ and $\mathcal{L}_{{\rm M}}$
describe the capacitive and inductive couplings between the components.

\subsubsection*{Lagrangian for resonator}

The resonator can be modeled as a distributed transmission line characterized
by inductance per unit length $L_{r}$ and capacitance per unit length
$C_{r}$~\cite{Pozar2021}. The resonant frequency of such a structure
is determined by the boundary conditions and the propagation characteristics
of electromagnetic waves in the waveguide. Let us consider a resonator
of length $\ell$ and open boundary conditions at both ends (for a
$\lambda/2$ resonator). The Lagrangian of the uncoupled resonator
is given by 
\begin{equation}
\mathcal{L}_{{\rm res}}=\int^{+\ell/2}_{-\ell/2}dx\left[\frac{C_{r}}{2}\left(\partial_{t}\Phi(x,t)\right)^{2}-\frac{1}{2L_{r}}\left(\partial_{x}\Phi(x,t)\right)^{2}\right],
\end{equation}
with boundary conditions imposed by the assumption of open-circuit
terminations at both ends: $\partial_{x}\Phi(\pm\ell/2,t)=0$, which
corresponds to vanishing current $I(\pm\ell/2,t)=\frac{1}{L_{r}}\partial_{x}\Phi(\pm\ell/2,t)$
at the endpoints.

We expand the flux field in the normal modes of the resonator, $f_{n}(x)$,
as 
\begin{equation}
\Phi(x,t)=\sum^{\infty}_{n=0}\phi_{n}(t)f_{n}(x),
\end{equation}
where $\phi_{n}(t)$ is the generalized coordinate of the $n$-th
mode. Due to the boundary conditions, $\partial_{x}f_{n}(\pm\ell/2)=0$,
the eigenmodes of the resonator take the form 
\begin{equation}
f_{n}(x)=\begin{cases}
\sqrt{\dfrac{2}{\ell}}\sin\left(\dfrac{n\pi x}{\ell}\right), & n=1,3,\dots\\
\sqrt{\dfrac{2}{\ell}}\cos\left(\dfrac{n\pi x}{\ell}\right), & n=2,4\dots
\end{cases}
\end{equation}
Substituting this expansion into the Lagrangian and integrating over
$x$, we obtain the mode Lagrangian for the resonator 
\begin{equation}
\mathcal{L}_{{\rm res}}=\sum_{n}\left[\frac{1}{2}C_{r}\dot{\phi}^{2}_{n}-\frac{1}{2}\frac{1}{L_{r}}k^{2}_{n}\phi^{2}_{n}\right],
\end{equation}
with $k_{n}=n\pi/\ell$, for $n\geq1$. The corresponding angular
frequency of the $n$-th mode is given by $\omega_{n}=v_{p}k_{n},$
where $v_{p}=1/\sqrt{L_{r}C_{r}}$ is the phase velocity in the resonator.
In the following, we consider the effective coupling of the qubit
to a single mode ($n=r$) of the resonator. Thus we can easily assume
$\Phi(x,t)=\phi_{r}(t)f_{r}(x)$.

\subsubsection*{Lagrangian for qubit}

The qubit’s degree of freedom is described by the node flux variable
$\phi_{q}(t)$. The qubit has a Lagrangian of the form 
\begin{equation}
\mathcal{L}_{{\rm qubit}}=\frac{C_{q}}{2}\dot{\phi}^{2}_{q}-\frac{1}{2L_{q}}\phi^{2}_{q}+E_{J}\cos\left(\frac{2\pi}{\Phi_{0}}\phi_{q}\right),
\end{equation}
where $C_{q}$ is the intrinsic qubit capacitance, $L_{q}$ is the
shunting inductance, $E_{J}$ is the Josephson energy.

\subsubsection*{Lagrangian for coupler}

The coupler is a tunable transmon, whose Josephson energy $E_{J,c}$
can be modulated by an external magnetic flux $\Phi_{{\rm ext}}$.
Its Lagrangian is given by 
\begin{equation}
\mathcal{L}_{{\rm coupler}}=\frac{C_{c}}{2}\dot{\phi}^{2}_{c}+E_{J,c}(\Phi_{{\rm ext}})\cos\left(\frac{2\pi}{\Phi_{0}}\phi_{c}\right).
\end{equation}
For analytical transparency, we model the coupler in the weakly anharmonic
limit as a single bosonic mode. The resulting expression captures
the leading flux dependence of the mediated interaction. A more complete
multilevel treatment would additionally generate small Kerr and dispersive
corrections, which do not alter the interference mechanism central
to this proposal.

\subsubsection*{Coupling Lagrangians}

The total capacitive coupling Lagrangian $\mathcal{L}_{{\rm C}}$
is the sum of three contributions: resonator-qubit and resonator-coupler
couplings via $C_{rq}$ and $C_{rc}$ at point $x_{C}$, respectively,
and qubit-coupler coupling via $C_{qc}$, and is given by 
\begin{align}
\mathcal{L}_{C}= & \frac{C_{rq}}{2}\left(\dot{\phi}_{q}-\partial_{t}\Phi(x_{C},t)\right)^{2}+\frac{C_{rc}}{2}(\dot{\phi}_{c}-\partial_{t}\Phi(x_{C},t))^{2}\nonumber \\
 & +\frac{C_{qc}}{2}(\dot{\phi}_{c}-\dot{\phi}_{q})^{2}.
\end{align}
The inductive coupling between the resonator and qubit is mediated
by a shared inductance $M$ at point $x_{M}$ 
\begin{align}
\mathcal{L}_{M} & =\frac{M}{L_{r}}I_{c}\sin(\frac{2\pi}{\Phi_{0}}\phi_{q})\partial_{x}\Phi(x_{M},t).
\end{align}

\subsection*{Circuit's Hamiltonian}

\subsubsection*{Legendre transformation}

To derive the Hamiltonian, we first identify the canonical momenta
$Q_{i}=\partial\mathcal{L}/\partial\dot{\phi}_{i}$ given by the vector
$\vec{Q}=\mathbf{C}\dot{\vec{\phi}}$, where $\vec{\phi}=(\Phi,\phi_{q},\phi_{c})^{T}$
and $\mathbf{C}$ is the symmetric $3\times3$ capacitance matrix
\begin{equation}
\mathbf{C}=\begin{pmatrix}C_{r}+C_{rq}+C_{rc} & -C_{rq} & -C_{rc}\\
-C_{rq} & C_{q}+C_{rq}+C_{qc} & -C_{qc}\\
-C_{rc} & -C_{qc} & C_{c}+C_{qc}+C_{rc}
\end{pmatrix}.
\end{equation}
We assume $C_{rq}\ll C_{qc},C_{rc}\ll C_{r},C_{q},C_{c}$ and obtain
the inverse of the capacitance matrix as
\begin{align}
\mathbf{C}^{-1} & =\frac{1}{\det(\mathbf{C})}\begin{pmatrix}C_{\Sigma q}C_{\Sigma c}-C^{2}_{qc} & C_{rq}C_{\Sigma c}+C_{qc}C_{rc} & C_{rc}C_{\Sigma q}+C_{rq}C_{qc}\\
C_{rq}C_{\Sigma c}+C_{qc}C_{rc} & C_{\Sigma r}C_{\Sigma c}-C^{2}_{rc} & C_{qc}C_{\Sigma r}+C_{rq}C_{rc}\\
C_{rc}C_{\Sigma q}+C_{rq}C_{qc} & C_{qc}C_{\Sigma r}+C_{rq}C_{rc} & C_{\Sigma r}C_{\Sigma q}-C^{2}_{rq}
\end{pmatrix}\nonumber \\
 & \approx\frac{1}{\det(\mathbf{C})}\begin{pmatrix}C_{\Sigma q}C_{\Sigma c} & C_{rq}C_{\Sigma c}+C_{qc}C_{rc} & C_{rc}C_{\Sigma q}+C_{rq}C_{qc}\\
C_{rq}C_{\Sigma c}+C_{qc}C_{rc} & C_{\Sigma r}C_{\Sigma c} & C_{qc}C_{\Sigma r}+C_{rq}C_{rc}\\
C_{rc}C_{\Sigma q}+C_{rq}C_{qc} & C_{qc}C_{\Sigma r}+C_{rq}C_{rc} & C_{\Sigma r}C_{\Sigma q}
\end{pmatrix},
\end{align}
where
\begin{align}
C_{\Sigma r} & =C_{r}+C_{rq}+C_{rc}\approx C_{r},\\
C_{\Sigma q} & =C_{q}+C_{rq}+C_{qc}\approx C_{q},\\
C_{\Sigma c} & =C_{c}+C_{qc}+C_{rc}\approx C_{c}.
\end{align}
Moreover, the determinant of the capacitance matrix is given by
\begin{align}
\det(\mathbf{C}) & =C_{\Sigma r}C_{\Sigma q}C_{\Sigma c}-C_{\Sigma r}C^{2}_{qc}-C_{\Sigma q}C^{2}_{rc}-C_{\Sigma c}C^{2}_{rq}-2C_{rq}C_{rc}C_{qc}\nonumber \\
 & \approx C_{\Sigma r}C_{\Sigma q}C_{\Sigma c}
\end{align}

The Hamiltonian of the coupled system is obtained via a Legendre transformation
of the total Lagrangian, $H=\sum_{i\in\{r,q,c\}}q_{i}\dot{\phi}_{i}-\mathcal{L}$.
The kinetic energy is expressed in terms of the conjugate charges
$\mathbf{q}=(q_{r},q_{q},q_{c})^{T}$ and the inverse of the capacitance
matrix $\mathbf{C}^{-1}$, yielding 
\begin{align}
H= & \frac{1}{2}\sum_{i,j\in\{r,q,c\}}(\mathbf{C}^{-1})_{ij}q_{i}q_{j}+\frac{k^{2}_{r}}{2L_{r}}\phi^{2}_{r}+\frac{1}{2L_{q}}\phi^{2}_{q}\nonumber \\
 & -E_{J}\cos\left(\frac{2\pi}{\Phi_{0}}\phi_{q}\right)-E_{J,c}(\Phi_{\text{ext}})\cos\left(\frac{2\pi}{\Phi_{0}}\phi_{c}\right)\nonumber \\
 & -\frac{M}{L_{r}}I_{c}\sin\left(\frac{2\pi}{\Phi_{0}}\phi_{q}\right)\partial_{x}f_{r}(x_{M})\phi_{r}.
\end{align}
Substituting the elements of $\mathbf{C}^{-1}$ in the above equation
and evaluating $\partial_{x}f_{r}(x)$ at the current antinode $x=x_{M}$, where it is maximal, we obtain
\begin{align}
H\approx & 4E_{C,r}n^{2}_{r}+4E_{C,q}n^{2}_{q}+4E_{C,c}n^{2}_{c}\nonumber \\
 & +4e^{2}\frac{C_{rq}C_{c}+C_{qc}C_{rc}}{C_{r}C_{q}C_{c}}n_{r}n_{q}+4e^{2}\left(\frac{C_{rc}}{C_{r}C_{c}}+\frac{C_{rq}C_{qc}}{C_{r}C_{q}C_{c}}\right)n_{r}n_{c}\nonumber \\
 & +4e^{2}\left(\frac{C_{qc}}{C_{q}C_{c}}+\frac{C_{rq}C_{rc}}{C_{r}C_{q}C_{c}}\right)n_{q}n_{c}\nonumber \\
 & +\frac{k^{2}_{r}}{2L_{r}}\phi^{2}_{r}+\frac{1}{2L_{q}}\phi^{2}_{q}\nonumber \\
 & -E_{J}\cos\left(\frac{2\pi}{\Phi_{0}}\phi_{q}\right)-E_{J,c}(\Phi_{\text{ext}})\cos\left(\frac{2\pi}{\Phi_{0}}\phi_{c}\right)\nonumber \\
 & -\frac{M}{L_{r}}I_{c}\sin\left(\frac{2\pi}{\Phi_{0}}\phi_{q}\right)\phi_{r},
\end{align}
where $n_{i}=\frac{q_{i}}{2e},$ represents the number of Cooper pairs
corresponding to the charge $q_{i}$, and $E_{C,i}=\frac{e^{2}}{2C_{i}},$
denotes the charging energy of the respective subsystem for $i\in\{r,q,c\}$.

\subsubsection*{Hamiltonian Quantization}

To transition from the classical Hamiltonian to its quantum mechanical
counterpart, we promote the canonical variables to quantum operators:
$\phi_{i}\rightarrow\hat{\phi}_{i}$ and $n_{i}\rightarrow\hat{n}_{i}$,
satisfying the commutation relation $[\hat{\phi}_{i},\hat{n}_{j}]=i\hbar/(2e)\delta_{ij}$.
We treat each subsystem according to its specific energy regime and
anharmonicity.

The resonator is modeled as a linear harmonic oscillator. We introduce
the bosonic creation and annihilation operators, $\hat{a}^{\dagger}$
and $\hat{a}$, such that the flux and charge operators are expressed
in terms of their zero-point fluctuations (ZPFs): 
\begin{align}
\hat{\phi}_{r} & =\phi_{\text{zpf},r}(\hat{a}+\hat{a}^{\dagger}),\\
\hat{n}_{r} & =in_{\text{zpf},r}(\hat{a}^{\dagger}-\hat{a}),
\end{align}
where $\phi_{\text{zpf},r}=\sqrt{\hbar Z_{r}/2}$ and $n_{\text{zpf},r}=\frac{1}{2e}\sqrt{\hbar/(2Z_{r})}$
with the characteristic impedance $Z_{r}=\sqrt{L_{r}/C_{r}}$. Substituting
these into the free resonator Hamiltonian yields the diagonal form
$H_{r}=\hbar\omega_{r}\hat{a}^{\dagger}\hat{a}$, with $\omega_{r}=1/\sqrt{L_{r}C_{r}}$.

The tunable coupler is assumed to operate in the transmon regime,
characterized by a large $E_{J,c}/E_{C,c}\gg1$ ratio. Consequently,
the phase fluctuations are small, allowing us to Taylor-expand the
Josephson cosine potential up to the fourth order: 
\begin{equation}
-E_{J,c}(\Phi_{\text{ext}})\cos\left(\frac{2\pi}{\Phi_{0}}\hat{\phi}_{c}\right)\approx-E_{J,c}+\frac{1}{2L_{J,c}}\hat{\phi}^{2}_{c}-\frac{E_{J,c}}{24}\left(\frac{2\pi}{\Phi_{0}}\hat{\phi}_{c}\right)^{4},
\end{equation}
where $L_{J,c}=(\Phi_{0}/2\pi)^{2}/E_{J,c}(\Phi_{\text{ext}})$. The
quadratic term combines with the capacitive energy to form a harmonic
oscillator, which we quantize using the operators $\hat{c}$ and $\hat{c}^{\dagger}$:
\begin{align}
\hat{\phi}_{c} & =\phi_{\text{zpf},c}(\hat{c}+\hat{c}^{\dagger}),\\
\hat{n}_{c} & =in_{\text{zpf},c}(\hat{c}^{\dagger}-\hat{c}),
\end{align}
with ZPFs defined similarly to the resonator using $C_{c}$ and $L_{Jc}$.
The coupler Hamiltonian then takes the form of a Duffing oscillator:
\begin{equation}
H_{c}=\hbar\omega_{c}\hat{c}^{\dagger}\hat{c}-\frac{\alpha_{c}}{2}\hat{c}^{\dagger}\hat{c}^{\dagger}\hat{c}\hat{c},
\end{equation}
where $\hbar\omega_{c}=\sqrt{8E_{J,c}E_{C,c}}-E_{C,c}$, and the anharmonicity
is governed by $\alpha_{c}=\frac{E_{J,c}}{2}\left(\frac{2\pi}{\Phi_{0}}\phi_{\text{zpf},c}\right)^{4}\approx E_{C,c}$.

The qubit is modeled as a fluxonium, featuring a large inductive shunt
($1/2L_{q}$) and an intermediate $E_{J}/E_{C,q}$ ratio. Due to its
strong anharmonicity and non-perturbative potential, it cannot be
approximated as a weakly nonlinear oscillator. Instead, we define
the bare fluxonium Hamiltonian
\begin{equation}
H_{q}=4E_{C,q}\hat{n}^{2}_{q}+\frac{1}{2L_{q}}\hat{\phi}^{2}_{q}-E_{J}\cos\left(\frac{2\pi}{\Phi_{0}}\hat{\phi}_{q}\right),\label{eq:H_bare_fluxonium}
\end{equation}
and solve the time-independent Schr\"odinger equation $H_{q}|k\rangle=E_{k}|k\rangle$
numerically using \textsc{scQubits} package~\cite{Groszkowski2021}.
Restricting the dynamics to the two lowest computational energy levels
($\{|0\rangle,|1\rangle\}$), the qubit Hamiltonian becomes
\begin{equation}
H_{q}=\frac{\hbar\omega_{q}}{2}\hat{\sigma}_{z},
\end{equation}
where $\hbar\omega_{q}=E_{1}-E_{0}$. The coupling operators must
be projected into this eigenbasis by computing their numerical matrix
elements. The charge operator and the nonlinear inductive coupling
term are expressed as
\begin{align}
\hat{n}_{q} & \approx\sum_{k,l\in\{0,1\}}\langle k|\hat{n}_{q}|l\rangle|k\rangle\langle l|=n_{01}\hat{\sigma}_{y},\\
\sin\left(\frac{2\pi}{\Phi_{0}}\hat{\phi}_{q}\right) & \approx\sum_{k,l\in\{0,1\}}\left\langle k|\sin\left(\frac{2\pi}{\Phi_{0}}\hat{\phi}_{q}\right)|l\right\rangle |k\rangle\langle l|=s_{01}\hat{\sigma}_{x},
\end{align}
where $n_{01}=\text{Im}[\langle0|\hat{n}_{q}|1\rangle]$ and $s_{01}=\text{Re}[\langle0|\sin(2\pi\hat{\phi}_{q}/\Phi_{0})|1\rangle]$
are numerically evaluated constants (see table~\eqref{tab:circuit_parameters-1}).
We note that the exact Pauli operator assignment depends on the chosen
gauge and parity of the eigenstates; here we assign $\hat{\sigma}_{y}=i(\hat{\sigma}^{-}-\hat{\sigma}^{+})$
to the charge matrix element assuming purely imaginary transition
elements, and $\hat{\sigma}_{x}=\hat{\sigma}^{-}+\hat{\sigma}^{+}$
to the sine term.

Substituting the quantized variables into the coupling terms of the
classical Hamiltonian, we derive the full quantum Hamiltonian. Defining
the effective coupling strengths
\begin{align}
\hbar g^{C}_{rq} & =4e^{2}\frac{C_{rq}C_{c}+C_{qc}C_{rc}}{C_{r}C_{q}C_{c}}n_{\text{zpf},r}n_{01},\\
\hbar g^{C}_{rc} & =4e^{2}\frac{C_{rc}C_{q}+C_{rq}C_{qc}}{C_{r}C_{q}C_{c}}n_{\text{zpf},r}n_{\text{zpf},c},\\
\hbar g^{C}_{qc} & =4e^{2}\frac{C_{qc}C_{r}+C_{rq}C_{rc}}{C_{r}C_{q}C_{c}}n_{\text{zpf},c}n_{01},\\
\hbar g^{L}_{rq} & =\frac{MI_{c}}{L_{r}}\phi_{\text{zpf},r}s_{01},
\end{align}
the complete quantized Hamiltonian is given by
\begin{align}
H= & \hbar\omega_{r}\hat{a}^{\dagger}\hat{a}+\hbar\omega_{c}\hat{c}^{\dagger}\hat{c}-\frac{\alpha_{c}}{2}\hat{c}^{\dagger}\hat{c}^{\dagger}\hat{c}\hat{c}+\frac{\hbar\omega_{q}}{2}\hat{\sigma}_{z}\nonumber \\
 & +i\hbar g^{C}_{rq}(\hat{a}^{\dagger}-\hat{a})\hat{\sigma}_{y}-\hbar g^{L}_{rq}(\hat{a}^{\dagger}+\hat{a})\hat{\sigma}_{x}\nonumber \\
 & -\hbar g^{C}_{rc}(\hat{a}^{\dagger}-\hat{a})(\hat{c}^{\dagger}-\hat{c})\nonumber \\
 & +i\hbar g^{C}_{qc}(\hat{c}^{\dagger}-\hat{c})\hat{\sigma}_{y}.\label{eq:full_quantized_H}
\end{align}

\subsubsection*{Qubit--Resonator effective Hamiltonian}

In our architecture, the tunable coupler is designed to mediate interactions
between the resonator and the qubit without exchanging real excitations.
To rigorously eliminate the coupler degrees of freedom from the dynamics,
we assume the coupler is in the dispersive regime with respect to
both the resonator and the qubit. This implies that the detunings
$\Delta_{rc}=\omega_{r}-\omega_{c}$ and $\Delta_{qc}=\omega_{q}-\omega_{c}$
are much larger than their respective coupling strengths, i.e., $|\Delta_{rc}|\gg|g^{C}_{rc}|$
and $|\Delta_{qc}|\gg|g^{C}_{qc}|$.

We apply a unitary SW transformation $U=e^{\hat{S}}$, where the anti-Hermitian
generator $\hat{S}=-\hat{S}^{\dagger}$ is chosen in perturbation
theory to eliminate the linear coupling terms involving the coupler.
The effective Hamiltonian is given by the Baker-Campbell-Hausdorff
expansion
\begin{equation}
H_{\text{eff}}=e^{\hat{S}}He^{-\hat{S}}\approx H_{0}+V_{rq}+V_{c}+[\hat{S},H_{0}]+[\hat{S},V_{rq}]+\frac{1}{2}[\hat{S},V_{c}],
\end{equation}
where $H_{0}$ encompasses the uncoupled Hamiltonian of all three
subsystems, $V_{rq}$ contains the direct resonator-qubit couplings,
and $V_{c}$ contains the coupler-dependent interaction terms. To
eliminate $V_{c}$ to first order, we require $[\hat{S},H_{0}]=-V_{c}$.

Given the interaction terms $V_{c}=-\hbar g^{C}_{rc}(\hat{a}^{\dagger}-\hat{a})(\hat{c}^{\dagger}-\hat{c})+i\hbar g^{C}_{qc}(\hat{c}^{\dagger}-\hat{c})\hat{\sigma}_{y}$,
the generator $\hat{S}$ takes the form $\hat{S}=\hat{S}_{rc}+\hat{S}_{qc}$,
with
\begin{align}
\hat{S}_{rc} & =g^{C}_{rc}\left(\frac{\hat{a}^{\dagger}\hat{c}-\hat{a}\hat{c}^{\dagger}}{\omega_{r}-\omega_{c}}+\frac{\hat{a}\hat{c}-\hat{a}^{\dagger}\hat{c}^{\dagger}}{\omega_{r}+\omega_{c}}\right),\\
\hat{S}_{qc} & =g^{C}_{qc}\left(\frac{\hat{c}^{\dagger}\hat{\sigma}^{-}-\hat{c}\hat{\sigma}^{+}}{\omega_{q}-\omega_{c}}+\frac{\hat{c}^{\dagger}\hat{\sigma}^{+}-\hat{c}\hat{\sigma}^{-}}{\omega_{q}+\omega_{c}}\right).
\end{align}

By substituting $\hat{S}$ back into the expansion the second-order
perturbation yields Lamb shifts to the bare frequencies (absorbed
into renormalized frequencies $\tilde{\omega}_{r}$ and $\tilde{\omega}_{q}$)
and an induced effective coupling between the resonator and the qubit.
Furthermore, assuming the tunable coupler remains in its ground state
($\langle\hat{c}^{\dagger}\hat{c}\rangle=0$), we project out the
coupler degrees of freedom, rendering its anharmonicity irrelevant
in the resulting second-order effective Hamiltonian and yields the
renormalized frequencies (absorbing the Lamb shifts) for the resonator
and the qubit
\begin{align}
\tilde{\omega}_{r} & =\omega_{r}+(g^{C}_{rc})^{2}\left(\frac{1}{\omega_{r}-\omega_{c}}-\frac{1}{\omega_{r}+\omega_{c}}\right),\\
\tilde{\omega}_{q} & =\omega_{q}+(g^{C}_{qc})^{2}\left(\frac{1}{\omega_{q}-\omega_{c}}+\frac{1}{\omega_{q}+\omega_{c}}\right).
\end{align}

The resulting effective Hamiltonian $H_{\text{eff}}$ acting strictly
on the resonator-qubit subspace is
\begin{align}
H_{\text{eff}}= & \hbar\tilde{\omega}_{r}\hat{a}^{\dagger}\hat{a}+\frac{\hbar\tilde{\omega}_{q}}{2}\hat{\sigma}_{z}\nonumber \\
 & +i\hbar\tilde{g}^{C}_{rq}(\hat{a}^{\dagger}-\hat{a})\hat{\sigma}_{y}-\hbar g^{L}_{rq}(\hat{a}^{\dagger}+\hat{a})\hat{\sigma}_{x},\label{eq:Heff_no_coupler}
\end{align}
where the renormalized capacitive coupling is $\tilde{g}^{C}_{rq}=g^{C}_{rq}+g^{C}_{\text{ind}}$.
The induced coupling strength $g^{C}_{\text{ind}}$ arises from the
cross-term $\frac{1}{2}([\hat{S}_{rc},V_{qc}]+[\hat{S}_{qc},V_{rc}])$.
Retaining the counter-rotating terms without making further approximations
at this stage yields the exact second-order expression 
\begin{equation}
g^{C}_{\text{ind}}=\frac{1}{2}g^{C}_{rc}g^{C}_{qc}\left(\frac{1}{\omega_{r}-\omega_{c}}-\frac{1}{\omega_{r}+\omega_{c}}+\frac{1}{\omega_{q}-\omega_{c}}-\frac{1}{\omega_{q}+\omega_{c}}\right).
\end{equation}

By tuning the external magnetic flux threading the coupler $\Phi_{\text{ext}}$,
one alters $E_{Jc}(\Phi_{\text{ext}})$, which modifies the coupler
frequency $\omega_{c}$ and its ZPFs, thereby providing \textit{in
situ} tunability of $\tilde{g}^{C}_{rq}$. This effective Hamiltonian
(Eq.~\ref{eq:Heff_no_coupler}) captures the full anisotropic physics
required for achieving dispersive shift cancellation, as the tunable
capacitive coupling $\tilde{g}^{C}_{rq}$ can be made to destructively
interfere with the static inductive coupling $g^{L}_{rq}$.

\subsubsection*{Dispersive Hamiltonian}

To derive the state-dependent dispersive Hamiltonian, we start with
the effective Hamiltonian describing the interaction between the resonator
and the qubit
\begin{equation}
H_{\text{eff}}=\hbar\tilde{\omega}_{r}\hat{a}^{\dagger}\hat{a}+\frac{\hbar\tilde{\omega}_{q}}{2}\hat{\sigma}_{z}+i\hbar\tilde{g}^{C}_{rq}(\hat{a}^{\dagger}-\hat{a})\hat{\sigma}_{y}-\hbar g^{L}_{rq}(\hat{a}^{\dagger}+\hat{a})\hat{\sigma}_{x}.\label{eq:H_eff}
\end{equation}
Expressing the Pauli operators in terms of raising and lowering operators,
$\hat{\sigma}_{x}=\hat{\sigma}_{+}+\hat{\sigma}_{-}$ and $\hat{\sigma}_{y}=i(\hat{\sigma}_{-}-\hat{\sigma}_{+})$,
the interaction term $V$ becomes
\begin{equation}
V=\hbar g_{{\rm JC}}(\hat{a}^{\dagger}\hat{\sigma}_{-}+\hat{a}\hat{\sigma}_{+})+\hbar g_{{\rm AJC}}(\hat{a}^{\dagger}\hat{\sigma}_{+}+\hat{a}\hat{\sigma}_{-}),\label{eq:V}
\end{equation}
where $g_{{\rm JC}}=-(\tilde{g}^{C}_{rq}+g^{L}_{rq})$ represents
the coupling strength of the co-rotating (JC) terms, and $g_{{\rm AJC}}=\tilde{g}^{C}_{rq}-g^{L}_{rq}$
represents the counter-rotating (AJC) terms.

To eliminate the off-diagonal interaction terms in the dispersive
regime, where detunings $\Delta=\tilde{\omega}_{q}-\tilde{\omega}_{r}$
and $\Sigma=\tilde{\omega}_{q}+\tilde{\omega}_{r}$ are much larger
than the coupling strengths ($|\Delta|,|\Sigma|\gg|g_{-}|,|g_{+}|$),
we apply a unitary SW transformation $U=\exp(\hat{S})$. The generator
$\hat{S}$ is chosen such that $[H_{0},\hat{S}]=-V$, yielding
\begin{equation}
\hat{S}=\frac{g_{{\rm JC}}}{\Delta}\hat{a}^{\dagger}\hat{\sigma}_{-}-\frac{g_{{\rm JC}}}{\Delta}\hat{a}\hat{\sigma}_{+}-\frac{g_{{\rm AJC}}}{\Sigma}\hat{a}^{\dagger}\hat{\sigma}_{+}+\frac{g_{{\rm AJC}}}{\Sigma}\hat{a}\hat{\sigma}_{-}.\label{eq:S_gen}
\end{equation}

Expanding the transformed Hamiltonian to second order in the coupling
strengths, $H_{\text{disp}}\approx H_{0}+\frac{1}{2}[S,V]$, we obtain
\begin{equation}
H_{\text{disp}}=\hbar\tilde{\omega}_{r}\hat{a}^{\dagger}\hat{a}+\frac{\hbar\tilde{\omega}_{q}}{2}\hat{\sigma}_{z}+\hbar\chi(\hat{a}^{\dagger}\hat{a}+\frac{1}{2})\hat{\sigma}_{z},\label{eq:H_disp_chi}
\end{equation}
where the state-dependent dispersive shift $\chi$ is given by the
contributions from both the JC and AJC interactions
\begin{equation}
\chi=\frac{g^{2}_{{\rm JC}}}{\Delta}+\frac{g^{2}_{{\rm JC}}}{\Sigma}.\label{eq:chi}
\end{equation}
By grouping the terms, the Hamiltonian can be written in a state-dependent
effective frequency form
\begin{equation}
H_{\text{disp}}=\hbar\left(\tilde{\omega}_{r}+\chi\hat{\sigma}_{z}\right)\hat{a}^{\dagger}\hat{a}+\frac{\hbar}{2}\left(\tilde{\omega}_{q}+\chi\right)\hat{\sigma}_{z}.\label{eq:H_disp_final}
\end{equation}
This explicitly demonstrates how the interplay between capacitive
and inductive couplings modifies the dispersive shift $\chi$, which
is the fundamental mechanism allowing for the cancellation of photon-shot-noise-induced
dephasing when $\chi\rightarrow0$.

\subsubsection*{Higher-Order Kerr and Dispersive Corrections}

In the main text, we numerically assess that the higher-order effects
do not distort the full dynamics across the entire parameter space,
by comparing the second-order analytical predictions directly against
the multi-level ED and non-perturbative L\"owdin downfolding across
the full coupler frequency sweep. The relative deviation $|g_{\text{ED}}-g_{\text{SW}}|/|g_{\text{max}}|$
is strictly bounded below $1.8\%$, and the shift in the predicted
sweet-spot frequency $\bar{\omega}_{c}$ is under $0.2\%$, confirming
that higher-order non-linearities leave the interference nodes and
coherence sweet spots intact.

\section*{Numerical Extraction of Effective Coupling Strengths}

In order to extract the effective couplings exactly from the diagonalized
numerical Hamiltonian, we employ a non-perturbative block-diagonalization
technique based on L\"owdin symmetric orthogonalization~\cite{Loewdin1950,Loewdin1962,Winkler2003}.

Let $H$ be the full system Hamiltonian acting on the total Hilbert
space, and $\{|\tilde{\psi}_{k}\rangle\}$ be its exact eigenvectors
with corresponding eigenvalues $E_{k}$. We define a target subspace
$\mathcal{P}$ spanned by a set of bare states $\{|\phi_{i}\rangle\}^{N}_{i=1}$,
where $N$ is the dimension of the subspace. Our goal is to construct
an effective $N\times N$ Hamiltonian, $H_{\text{eff}}$, acting only
within $\mathcal{P}$, whose eigenvalues exactly match the $N$ selected
dressed eigenvalues of the full Hamiltonian, and whose eigenvectors
maintain maximal overlap with the bare states.

We first identify the $N$ exact eigenstates $\{|\tilde{\psi}_{m}\rangle\}^{N}_{m=1}$
that have the highest projection onto our chosen bare states $\{|\phi_{i}\rangle\}$.
We construct the projection matrix $V$ with elements defined by the
overlaps
\begin{equation}
V_{im}=\langle\phi_{i}|\tilde{\psi}_{m}\rangle.
\end{equation}
Because the chosen exact eigenstates have components outside the target
subspace $\mathcal{P}$, the projected vectors are not strictly orthogonal.
The overlap matrix of these projected states is given by 
\begin{equation}
S=VV^{\dagger}.
\end{equation}

To restore orthonormality while minimizing the distance between the
original and the new basis (in a least-squares sense), we apply the
L\"owdin symmetric orthogonalization. The transformation matrix is
$S^{-1/2}$, leading to the effective Hamiltonian
\begin{equation}
H_{\text{eff}}=S^{-1/2}V\Lambda V^{\dagger}S^{-1/2},\label{eq:Heff_general}
\end{equation}
where $\Lambda=\text{diag}(E_{1},E_{2},\dots,E_{N})$ is the diagonal
matrix of the selected exact eigenvalues. The off-diagonal elements
of $H_{\text{eff}}$ directly yield the exact effective couplings
between the basis states of $\mathcal{P}$, including all higher-order
dispersive shifts and virtual transitions.

The JC coupling governs the coherent energy exchange between the qubit
and the resonator, corresponding to transitions between $|e,0\rangle$
and $|g,1\rangle$. Since no other bare states cross this subspace
in the dispersive or moderately resonant regimes, a standard two-level
projection is sufficient.

We define the bare subspace as $\mathcal{P}_{\text{JC}}=\{|\phi_{1}\rangle=|e,0\rangle,|\phi_{2}\rangle=|g,1\rangle\}$.
Following Eq.~\ref{eq:Heff_general}, we construct the $2\times2$
effective Hamiltonian
\begin{equation}
H^{(\text{JC})}_{\text{eff}}=\begin{pmatrix}\tilde{\omega}_{e0} & g_{\text{JC}}\\
g^{*}_{\text{JC}} & \tilde{\omega}_{g1}
\end{pmatrix}.
\end{equation}
The off-diagonal element gives the exact effective JC coupling, $g_{\text{JC}}=\langle e,0|H^{(\text{JC})}_{\text{eff}}|g,1\rangle$,
while the diagonal elements provide the fully dressed energies, implicitly
containing the Lamb shifts and dispersive shifts induced by the coupler.

The AJC coupling describes the counter-rotating two-excitation generation/annihilation
process between $|g,0\rangle$ and $|e,1\rangle$. A naive extraction
would use a $2\times2$ subspace $\mathcal{P}_{\text{AJC}}=\{|g,0\rangle,|e,1\rangle\}$.
However, near the resonance $\omega_{q}\approx\omega_{r}$, the energy
of the $|e,1\rangle$ state ($\approx\omega_{q}+\omega_{r}$) becomes
nearly degenerate with the $|g,2\rangle$ state ($\approx2\omega_{r}$).

In perturbative treatments, the coupler mediates an interaction between
$|e,1\rangle$ and $|g,2\rangle$. If $|g,2\rangle$ is treated as
an out-of-subspace state and integrated out, the energy denominator
$(E_{e1}-E_{g2})$ vanishes at resonance, leading to an unphysical
divergence in the calculated $g_{\text{AJC}}$.

To eliminate this singularity and capture the correct physics, we
extend the L\"owdin projection method to a $3\times3$ subspace that
explicitly includes the intruder state. We define the extended target
subspace as
\begin{equation}
\mathcal{P}_{\text{ext}}=\{|\phi_{1}\rangle=|g,0\rangle,|\phi_{2}\rangle=|e,1\rangle,|\phi_{3}\rangle=|g,2\rangle\}.
\end{equation}

By selecting the three dressed states corresponding to these bare
states and applying the transformation in Eq.~\ref{eq:Heff_general},
we obtain a $3\times3$ effective Hamiltonian
\begin{equation}
H^{(\text{ext})}_{\text{eff}}=\begin{pmatrix}\tilde{\omega}_{g0} & g_{\text{AJC}} & \Omega_{02}\\
g^{*}_{\text{AJC}} & \tilde{\omega}_{e1} & J_{12}\\
\Omega^{*}_{02} & J^{*}_{12} & \tilde{\omega}_{g2}
\end{pmatrix}.
\end{equation}
Because the resonant interaction between $|e,1\rangle$ and $|g,2\rangle$
is now treated exactly within the effective block (via the matrix
element $J_{12}$), the singularity is strictly avoided. The effective
AJC coupling is then safely and accurately extracted as
\begin{equation}
g_{\text{AJC}}=\langle g,0|H^{(\text{ext})}_{\text{eff}}|e,1\rangle.
\end{equation}

\bibliography{Refs}

\begin{thebibliography}{10}
\urlstyle{rm}
\expandafter\ifx\csname url\endcsname\relax
  \def\url#1{\texttt{#1}}\fi
\expandafter\ifx\csname urlprefix\endcsname\relax\def\urlprefix{URL }\fi
\expandafter\ifx\csname doiprefix\endcsname\relax\def\doiprefix{DOI: }\fi
\providecommand{\bibinfo}[2]{#2}
\providecommand{\eprint}[2][]{\url{#2}}

\bibitem{Blais2021}
\bibinfo{author}{Blais, A.}, \bibinfo{author}{Grimsmo, A.~L.},
  \bibinfo{author}{Girvin, S.~M.} \& \bibinfo{author}{Wallraff, A.}
\newblock \bibinfo{journal}{\bibinfo{title}{Circuit quantum electrodynamics}}.
\newblock {\emph{\JournalTitle{Rev. Mod. Phys.}}}
  \textbf{\bibinfo{volume}{93}}, \bibinfo{pages}{025005},
  \doiprefix\url{10.1103/RevModPhys.93.025005} (\bibinfo{year}{2021}).

\bibitem{Schoelkopf2008}
\bibinfo{author}{Schoelkopf, R.~J.} \& \bibinfo{author}{Girvin, S.~M.}
\newblock \bibinfo{journal}{\bibinfo{title}{Wiring up quantum systems}}.
\newblock {\emph{\JournalTitle{Nature}}} \textbf{\bibinfo{volume}{451}},
  \bibinfo{pages}{664--669}, \doiprefix\url{10.1038/451664a}
  (\bibinfo{year}{2008}).

\bibitem{Wallraff2004}
\bibinfo{author}{Wallraff, A.} \emph{et~al.}
\newblock \bibinfo{journal}{\bibinfo{title}{Strong coupling of a single photon
  to a superconducting qubit using circuit quantum electrodynamics}}.
\newblock {\emph{\JournalTitle{Nature}}} \textbf{\bibinfo{volume}{431}},
  \bibinfo{pages}{162--167}, \doiprefix\url{10.1038/nature02851}
  (\bibinfo{year}{2004}).

\bibitem{Blais2004}
\bibinfo{author}{Blais, A.}, \bibinfo{author}{Huang, R.-S.},
  \bibinfo{author}{Wallraff, A.}, \bibinfo{author}{Girvin, S.~M.} \&
  \bibinfo{author}{Schoelkopf, R.~J.}
\newblock \bibinfo{journal}{\bibinfo{title}{Cavity quantum electrodynamics for
  superconducting electrical circuits: An architecture for quantum
  computation}}.
\newblock {\emph{\JournalTitle{Physical Review A}}}
  \textbf{\bibinfo{volume}{69}}, \bibinfo{pages}{062320},
  \doiprefix\url{10.1103/physreva.69.062320} (\bibinfo{year}{2004}).

\bibitem{Rabi1937}
\bibinfo{author}{Rabi, I.~I.}
\newblock \bibinfo{journal}{\bibinfo{title}{Space quantization in a gyrating
  magnetic field}}.
\newblock {\emph{\JournalTitle{Physical Review}}}
  \textbf{\bibinfo{volume}{51}}, \bibinfo{pages}{652--654},
  \doiprefix\url{10.1103/physrev.51.652} (\bibinfo{year}{1937}).

\bibitem{BocanegraGaray2024}
\bibinfo{author}{Bocanegra-Garay, I.~A.}, \bibinfo{author}{Castillo-Celeita,
  M.}, \bibinfo{author}{Negro, J.}, \bibinfo{author}{Nieto, L.~M.} \&
  \bibinfo{author}{G\'omez-Ruiz, F.~J.}
\newblock \bibinfo{journal}{\bibinfo{title}{Exploring supersymmetry:
  Interchangeability between jaynes-cummings and anti-jaynes-cummings models}}.
\newblock {\emph{\JournalTitle{Phys. Rev. Res.}}} \textbf{\bibinfo{volume}{6}},
  \bibinfo{pages}{043218}, \doiprefix\url{10.1103/PhysRevResearch.6.043218}
  (\bibinfo{year}{2024}).

\bibitem{Braak2011}
\bibinfo{author}{Braak, D.}
\newblock \bibinfo{journal}{\bibinfo{title}{Integrability of the rabi model}}.
\newblock {\emph{\JournalTitle{Phys. Rev. Lett.}}}
  \textbf{\bibinfo{volume}{107}}, \bibinfo{pages}{100401},
  \doiprefix\url{10.1103/PhysRevLett.107.100401} (\bibinfo{year}{2011}).

\bibitem{Baksic2014}
\bibinfo{author}{Baksic, A.} \& \bibinfo{author}{Ciuti, C.}
\newblock \bibinfo{journal}{\bibinfo{title}{Controlling discrete and continuous
  symmetries in “superradiant” phase transitions with circuit qed
  systems}}.
\newblock {\emph{\JournalTitle{Physical Review Letters}}}
  \textbf{\bibinfo{volume}{112}}, \bibinfo{pages}{173601},
  \doiprefix\url{10.1103/physrevlett.112.173601} (\bibinfo{year}{2014}).

\bibitem{Jaynes1963}
\bibinfo{author}{Jaynes, E.} \& \bibinfo{author}{Cummings, F.}
\newblock \bibinfo{journal}{\bibinfo{title}{Comparison of quantum and
  semiclassical radiation theories with application to the beam maser}}.
\newblock {\emph{\JournalTitle{Proceedings of the IEEE}}}
  \textbf{\bibinfo{volume}{51}}, \bibinfo{pages}{89--109},
  \doiprefix\url{10.1109/proc.1963.1664} (\bibinfo{year}{1963}).

\bibitem{Hofheinz2008}
\bibinfo{author}{Hofheinz, M.} \emph{et~al.}
\newblock \bibinfo{journal}{\bibinfo{title}{Generation of fock states in a
  superconducting quantum circuit}}.
\newblock {\emph{\JournalTitle{Nature}}} \textbf{\bibinfo{volume}{454}},
  \bibinfo{pages}{310--314}, \doiprefix\url{10.1038/nature07136}
  (\bibinfo{year}{2008}).

\bibitem{Deppe2008}
\bibinfo{author}{Deppe, F.} \emph{et~al.}
\newblock \bibinfo{journal}{\bibinfo{title}{Two-photon probe of the
  jaynes–cummings model and controlled symmetry breaking in circuit qed}}.
\newblock {\emph{\JournalTitle{Nature Physics}}} \textbf{\bibinfo{volume}{4}},
  \bibinfo{pages}{686--691}, \doiprefix\url{10.1038/nphys1016}
  (\bibinfo{year}{2008}).

\bibitem{Majer2007}
\bibinfo{author}{Majer, J.} \emph{et~al.}
\newblock \bibinfo{journal}{\bibinfo{title}{Coupling superconducting qubits via
  a cavity bus}}.
\newblock {\emph{\JournalTitle{Nature}}} \textbf{\bibinfo{volume}{449}},
  \bibinfo{pages}{443--447}, \doiprefix\url{10.1038/nature06184}
  (\bibinfo{year}{2007}).

\bibitem{Sillanpaeae2007}
\bibinfo{author}{Sillanp\"{a}\"{a}, M.~A.}, \bibinfo{author}{Park, J.~I.} \&
  \bibinfo{author}{Simmonds, R.~W.}
\newblock \bibinfo{journal}{\bibinfo{title}{Coherent quantum state storage and
  transfer between two phase qubits via a resonant cavity}}.
\newblock {\emph{\JournalTitle{Nature}}} \textbf{\bibinfo{volume}{449}},
  \bibinfo{pages}{438--442}, \doiprefix\url{10.1038/nature06124}
  (\bibinfo{year}{2007}).

\bibitem{Niemczyk2010}
\bibinfo{author}{Niemczyk, T.} \emph{et~al.}
\newblock \bibinfo{journal}{\bibinfo{title}{Circuit quantum electrodynamics in
  the ultrastrong-coupling regime}}.
\newblock {\emph{\JournalTitle{Nature Physics}}} \textbf{\bibinfo{volume}{6}},
  \bibinfo{pages}{772--776}, \doiprefix\url{10.1038/nphys1730}
  (\bibinfo{year}{2010}).

\bibitem{Casanova2010}
\bibinfo{author}{Casanova, J.}, \bibinfo{author}{Romero, G.},
  \bibinfo{author}{Lizuain, I.}, \bibinfo{author}{Garc\'{i}a-Ripoll, J.~J.} \&
  \bibinfo{author}{Solano, E.}
\newblock \bibinfo{journal}{\bibinfo{title}{Deep strong coupling regime of the
  jaynes-cummings model}}.
\newblock {\emph{\JournalTitle{Physical Review Letters}}}
  \textbf{\bibinfo{volume}{105}}, \bibinfo{pages}{263603},
  \doiprefix\url{10.1103/physrevlett.105.263603} (\bibinfo{year}{2010}).

\bibitem{TwyeffortIrish2007}
\bibinfo{author}{Twyeffort~Irish, E.~K.}
\newblock \bibinfo{journal}{\bibinfo{title}{Generalized rotating-wave
  approximation for arbitrarily large coupling}}.
\newblock {\emph{\JournalTitle{Physical Review Letters}}}
  \textbf{\bibinfo{volume}{99}}, \bibinfo{pages}{173601},
  \doiprefix\url{10.1103/physrevlett.99.173601} (\bibinfo{year}{2007}).

\bibitem{Zueco2009}
\bibinfo{author}{Zueco, D.}, \bibinfo{author}{Reuther, G.~M.},
  \bibinfo{author}{Kohler, S.} \& \bibinfo{author}{H\"{a}nggi, P.}
\newblock \bibinfo{journal}{\bibinfo{title}{Qubit-oscillator dynamics in the
  dispersive regime: Analytical theory beyond the rotating-wave
  approximation}}.
\newblock {\emph{\JournalTitle{Physical Review A}}}
  \textbf{\bibinfo{volume}{80}}, \bibinfo{pages}{033846},
  \doiprefix\url{10.1103/physreva.80.033846} (\bibinfo{year}{2009}).

\bibitem{Bloch1940}
\bibinfo{author}{Bloch, F.} \& \bibinfo{author}{Siegert, A.}
\newblock \bibinfo{journal}{\bibinfo{title}{Magnetic resonance for nonrotating
  fields}}.
\newblock {\emph{\JournalTitle{Physical Review}}}
  \textbf{\bibinfo{volume}{57}}, \bibinfo{pages}{522--527},
  \doiprefix\url{10.1103/physrev.57.522} (\bibinfo{year}{1940}).

\bibitem{Tuorila2010}
\bibinfo{author}{Tuorila, J.} \emph{et~al.}
\newblock \bibinfo{journal}{\bibinfo{title}{Stark effect and generalized
  bloch-siegert shift in a strongly driven two-level system}}.
\newblock {\emph{\JournalTitle{Physical Review Letters}}}
  \textbf{\bibinfo{volume}{105}}, \bibinfo{pages}{257003},
  \doiprefix\url{10.1103/physrevlett.105.257003} (\bibinfo{year}{2010}).

\bibitem{FornDiaz2010}
\bibinfo{author}{Forn-D\'{i}az, P.} \emph{et~al.}
\newblock \bibinfo{journal}{\bibinfo{title}{Observation of the bloch-siegert
  shift in a qubit-oscillator system in the ultrastrong coupling regime}}.
\newblock {\emph{\JournalTitle{Physical Review Letters}}}
  \textbf{\bibinfo{volume}{105}}, \bibinfo{pages}{237001},
  \doiprefix\url{10.1103/physrevlett.105.237001} (\bibinfo{year}{2010}).

\bibitem{Solano2003}
\bibinfo{author}{Solano, E.}, \bibinfo{author}{Agarwal, G.~S.} \&
  \bibinfo{author}{Walther, H.}
\newblock \bibinfo{journal}{\bibinfo{title}{Strong-driving-assisted
  multipartite entanglement in cavity qed}}.
\newblock {\emph{\JournalTitle{Physical Review Letters}}}
  \textbf{\bibinfo{volume}{90}}, \bibinfo{pages}{027903},
  \doiprefix\url{10.1103/physrevlett.90.027903} (\bibinfo{year}{2003}).

\bibitem{Beaudoin2011}
\bibinfo{author}{Beaudoin, F.}, \bibinfo{author}{Gambetta, J.~M.} \&
  \bibinfo{author}{Blais, A.}
\newblock \bibinfo{journal}{\bibinfo{title}{Dissipation and ultrastrong
  coupling in circuit qed}}.
\newblock {\emph{\JournalTitle{Phys. Rev. A}}} \textbf{\bibinfo{volume}{84}},
  \bibinfo{pages}{043832}, \doiprefix\url{10.1103/PhysRevA.84.043832}
  (\bibinfo{year}{2011}).

\bibitem{Ma2015}
\bibinfo{author}{Ma, K. K.~W.} \& \bibinfo{author}{Law, C.~K.}
\newblock \bibinfo{journal}{\bibinfo{title}{Three-photon resonance and
  adiabatic passage in the large-detuning rabi model}}.
\newblock {\emph{\JournalTitle{Physical Review A}}}
  \textbf{\bibinfo{volume}{92}}, \bibinfo{pages}{023842},
  \doiprefix\url{10.1103/physreva.92.023842} (\bibinfo{year}{2015}).

\bibitem{Ciuti2006}
\bibinfo{author}{Ciuti, C.} \& \bibinfo{author}{Carusotto, I.}
\newblock \bibinfo{journal}{\bibinfo{title}{Input-output theory of cavities in
  the ultrastrong coupling regime: The case of time-independent cavity
  parameters}}.
\newblock {\emph{\JournalTitle{Physical Review A}}}
  \textbf{\bibinfo{volume}{74}}, \bibinfo{pages}{033811},
  \doiprefix\url{10.1103/physreva.74.033811} (\bibinfo{year}{2006}).

\bibitem{Ridolfo2012}
\bibinfo{author}{Ridolfo, A.}, \bibinfo{author}{Leib, M.},
  \bibinfo{author}{Savasta, S.} \& \bibinfo{author}{Hartmann, M.~J.}
\newblock \bibinfo{journal}{\bibinfo{title}{Photon blockade in the ultrastrong
  coupling regime}}.
\newblock {\emph{\JournalTitle{Physical Review Letters}}}
  \textbf{\bibinfo{volume}{109}}, \bibinfo{pages}{193602},
  \doiprefix\url{10.1103/physrevlett.109.193602} (\bibinfo{year}{2012}).

\bibitem{Shang2019}
\bibinfo{author}{Shang, C.}, \bibinfo{author}{Shen, H.~Z.} \&
  \bibinfo{author}{Yi, X.~X.}
\newblock \bibinfo{journal}{\bibinfo{title}{Nonreciprocity in a strongly
  coupled three-mode optomechanical circulatory system}}.
\newblock {\emph{\JournalTitle{Optics Express}}} \textbf{\bibinfo{volume}{27}},
  \bibinfo{pages}{25882}, \doiprefix\url{10.1364/oe.27.025882}
  (\bibinfo{year}{2019}).

\bibitem{Mercurio2022}
\bibinfo{author}{Mercurio, A.} \emph{et~al.}
\newblock \bibinfo{journal}{\bibinfo{title}{Regimes of cavity qed under
  incoherent excitation: From weak to deep strong coupling}}.
\newblock {\emph{\JournalTitle{Physical Review Research}}}
  \textbf{\bibinfo{volume}{4}}, \bibinfo{pages}{023048},
  \doiprefix\url{10.1103/physrevresearch.4.023048} (\bibinfo{year}{2022}).

\bibitem{Shang2026}
\bibinfo{author}{Shang, C.} \& \bibinfo{author}{Shen, H.~Z.}
\newblock \bibinfo{journal}{\bibinfo{title}{Coupling enhancement and
  symmetrization in dissipative optomechanical systems}}.
\newblock {\emph{\JournalTitle{Annalen der Physik}}}
  \textbf{\bibinfo{volume}{538}}, \bibinfo{pages}{e00606},
  \doiprefix\url{10.1002/andp.202500606} (\bibinfo{year}{2026}).

\bibitem{Hwang2015}
\bibinfo{author}{Hwang, M.-J.}, \bibinfo{author}{Puebla, R.} \&
  \bibinfo{author}{Plenio, M.~B.}
\newblock \bibinfo{journal}{\bibinfo{title}{Quantum phase transition and
  universal dynamics in the rabi model}}.
\newblock {\emph{\JournalTitle{Physical Review Letters}}}
  \textbf{\bibinfo{volume}{115}}, \bibinfo{pages}{180404},
  \doiprefix\url{10.1103/physrevlett.115.180404} (\bibinfo{year}{2015}).

\bibitem{Hwang2016}
\bibinfo{author}{Hwang, M.-J.}, \bibinfo{author}{Kim, M.} \&
  \bibinfo{author}{Choi, M.-S.}
\newblock \bibinfo{journal}{\bibinfo{title}{Recurrent delocalization and
  quasiequilibration of photons in coupled systems in circuit quantum
  electrodynamics}}.
\newblock {\emph{\JournalTitle{Physical Review Letters}}}
  \textbf{\bibinfo{volume}{116}}, \bibinfo{pages}{153601},
  \doiprefix\url{10.1103/physrevlett.116.153601} (\bibinfo{year}{2016}).

\bibitem{LeBoite2016}
\bibinfo{author}{Le~Boit\'{e}, A.}, \bibinfo{author}{Hwang, M.-J.},
  \bibinfo{author}{Nha, H.} \& \bibinfo{author}{Plenio, M.~B.}
\newblock \bibinfo{journal}{\bibinfo{title}{Fate of photon blockade in the deep
  strong-coupling regime}}.
\newblock {\emph{\JournalTitle{Physical Review A}}}
  \textbf{\bibinfo{volume}{94}}, \bibinfo{pages}{033827},
  \doiprefix\url{10.1103/physreva.94.033827} (\bibinfo{year}{2016}).

\bibitem{Shen2018}
\bibinfo{author}{Shen, H.~Z.}, \bibinfo{author}{Shang, C.},
  \bibinfo{author}{Zhou, Y.~H.} \& \bibinfo{author}{Yi, X.~X.}
\newblock \bibinfo{journal}{\bibinfo{title}{Unconventional single-photon
  blockade in non-markovian systems}}.
\newblock {\emph{\JournalTitle{Physical Review A}}}
  \textbf{\bibinfo{volume}{98}}, \bibinfo{pages}{023856},
  \doiprefix\url{10.1103/physreva.98.023856} (\bibinfo{year}{2018}).

\bibitem{Trivedi2019}
\bibinfo{author}{Trivedi, R.}, \bibinfo{author}{Radulaski, M.},
  \bibinfo{author}{Fischer, K.~A.}, \bibinfo{author}{Fan, S.} \&
  \bibinfo{author}{Vučković, J.}
\newblock \bibinfo{journal}{\bibinfo{title}{Photon blockade in weakly driven
  cavity quantum electrodynamics systems with many emitters}}.
\newblock {\emph{\JournalTitle{Physical Review Letters}}}
  \textbf{\bibinfo{volume}{122}}, \bibinfo{pages}{243602},
  \doiprefix\url{10.1103/physrevlett.122.243602} (\bibinfo{year}{2019}).

\bibitem{Lu2025}
\bibinfo{author}{Lu, Z.-G.}, \bibinfo{author}{Tian, G.}, \bibinfo{author}{Lü,
  X.-Y.} \& \bibinfo{author}{Shang, C.}
\newblock \bibinfo{journal}{\bibinfo{title}{Topological quantum batteries}}.
\newblock {\emph{\JournalTitle{Physical Review Letters}}}
  \textbf{\bibinfo{volume}{134}}, \bibinfo{pages}{180401},
  \doiprefix\url{10.1103/physrevlett.134.180401} (\bibinfo{year}{2025}).

\bibitem{Zheng2008}
\bibinfo{author}{Zheng, H.}, \bibinfo{author}{Zhu, S.~Y.} \&
  \bibinfo{author}{Zubairy, M.~S.}
\newblock \bibinfo{journal}{\bibinfo{title}{Quantum zeno and anti-zeno effects:
  Without the rotating-wave approximation}}.
\newblock {\emph{\JournalTitle{Phys. Rev. Lett.}}}
  \textbf{\bibinfo{volume}{101}}, \bibinfo{pages}{200404},
  \doiprefix\url{10.1103/PhysRevLett.101.200404} (\bibinfo{year}{2008}).

\bibitem{Ficek2002}
\bibinfo{author}{Ficek, Z.} \& \bibinfo{author}{Tana\'{s}, R.}
\newblock \bibinfo{journal}{\bibinfo{title}{Entangled states and collective
  nonclassical effects in two-atom systems}}.
\newblock {\emph{\JournalTitle{Physics Reports}}}
  \textbf{\bibinfo{volume}{372}}, \bibinfo{pages}{369--443},
  \doiprefix\url{10.1016/s0370-1573(02)00368-x} (\bibinfo{year}{2002}).

\bibitem{Jing2009}
\bibinfo{author}{Jing, J.}, \bibinfo{author}{Lü, Z.-G.} \&
  \bibinfo{author}{Ficek, Z.}
\newblock \bibinfo{journal}{\bibinfo{title}{Breakdown of the rotating-wave
  approximation in the description of entanglement of spin-anticorrelated
  states}}.
\newblock {\emph{\JournalTitle{Physical Review A}}}
  \textbf{\bibinfo{volume}{79}}, \bibinfo{pages}{044305},
  \doiprefix\url{10.1103/physreva.79.044305} (\bibinfo{year}{2009}).

\bibitem{Chen2010}
\bibinfo{author}{Chen, Q.-H.}, \bibinfo{author}{Yang, Y.},
  \bibinfo{author}{Liu, T.} \& \bibinfo{author}{Wang, K.-L.}
\newblock \bibinfo{journal}{\bibinfo{title}{Entanglement dynamics of two
  independent jaynes-cummings atoms without the rotating-wave approximation}}.
\newblock {\emph{\JournalTitle{Physical Review A}}}
  \textbf{\bibinfo{volume}{82}}, \bibinfo{pages}{052306},
  \doiprefix\url{10.1103/physreva.82.052306} (\bibinfo{year}{2010}).

\bibitem{Shang2024}
\bibinfo{author}{Shang, C.} \& \bibinfo{author}{Li, H.}
\newblock \bibinfo{journal}{\bibinfo{title}{Resonance-dominant optomechanical
  entanglement in open quantum systems}}.
\newblock {\emph{\JournalTitle{Physical Review Applied}}}
  \textbf{\bibinfo{volume}{21}}, \bibinfo{pages}{044048},
  \doiprefix\url{10.1103/physrevapplied.21.044048} (\bibinfo{year}{2024}).

\bibitem{Lue2012}
\bibinfo{author}{L\"{u}, Z.} \& \bibinfo{author}{Zheng, H.}
\newblock \bibinfo{journal}{\bibinfo{title}{Effects of counter-rotating
  interaction on driven tunneling dynamics: Coherent destruction of tunneling
  and bloch-siegert shift}}.
\newblock {\emph{\JournalTitle{Physical Review A}}}
  \textbf{\bibinfo{volume}{86}}, \bibinfo{pages}{023831},
  \doiprefix\url{10.1103/physreva.86.023831} (\bibinfo{year}{2012}).

\bibitem{Grimsmo2013}
\bibinfo{author}{Grimsmo, A.~L.} \& \bibinfo{author}{Parkins, S.}
\newblock \bibinfo{journal}{\bibinfo{title}{Cavity-qed simulation of
  qubit-oscillator dynamics in the ultrastrong-coupling regime}}.
\newblock {\emph{\JournalTitle{Phys. Rev. A}}} \textbf{\bibinfo{volume}{87}},
  \bibinfo{pages}{033814}, \doiprefix\url{10.1103/PhysRevA.87.033814}
  (\bibinfo{year}{2013}).

\bibitem{Ashhab2007}
\bibinfo{author}{Ashhab, S.}, \bibinfo{author}{Johansson, J.~R.},
  \bibinfo{author}{Zagoskin, A.~M.} \& \bibinfo{author}{Nori, F.}
\newblock \bibinfo{journal}{\bibinfo{title}{Two-level systems driven by
  large-amplitude fields}}.
\newblock {\emph{\JournalTitle{Phys. Rev. A}}} \textbf{\bibinfo{volume}{75}},
  \bibinfo{pages}{063414}, \doiprefix\url{10.1103/PhysRevA.75.063414}
  (\bibinfo{year}{2007}).

\bibitem{Ballester2012}
\bibinfo{author}{Ballester, D.}, \bibinfo{author}{Romero, G.},
  \bibinfo{author}{Garc\'{i}a-Ripoll, J.~J.}, \bibinfo{author}{Deppe, F.} \&
  \bibinfo{author}{Solano, E.}
\newblock \bibinfo{journal}{\bibinfo{title}{Quantum simulation of the
  ultrastrong-coupling dynamics in circuit quantum electrodynamics}}.
\newblock {\emph{\JournalTitle{Physical Review X}}}
  \textbf{\bibinfo{volume}{2}}, \bibinfo{pages}{021007},
  \doiprefix\url{10.1103/physrevx.2.021007} (\bibinfo{year}{2012}).

\bibitem{Deng2015}
\bibinfo{author}{Deng, C.}, \bibinfo{author}{Orgiazzi, J.-L.},
  \bibinfo{author}{Shen, F.}, \bibinfo{author}{Ashhab, S.} \&
  \bibinfo{author}{Lupascu, A.}
\newblock \bibinfo{journal}{\bibinfo{title}{Observation of floquet states in a
  strongly driven artificial atom}}.
\newblock {\emph{\JournalTitle{Physical Review Letters}}}
  \textbf{\bibinfo{volume}{115}}, \bibinfo{pages}{133601},
  \doiprefix\url{10.1103/physrevlett.115.133601} (\bibinfo{year}{2015}).

\bibitem{Yoshihara2017}
\bibinfo{author}{Yoshihara, F.} \emph{et~al.}
\newblock \bibinfo{journal}{\bibinfo{title}{Superconducting qubit–oscillator
  circuit beyond the ultrastrong-coupling regime}}.
\newblock {\emph{\JournalTitle{Nature Physics}}} \textbf{\bibinfo{volume}{13}},
  \bibinfo{pages}{44--47}, \doiprefix\url{10.1038/nphys3906}
  (\bibinfo{year}{2017}).

\bibitem{Braumueller2017}
\bibinfo{author}{Braumüller, J.} \emph{et~al.}
\newblock \bibinfo{journal}{\bibinfo{title}{Analog quantum simulation of the
  rabi model in the ultra-strong coupling regime}}.
\newblock {\emph{\JournalTitle{Nature Communications}}}
  \textbf{\bibinfo{volume}{8}}, \bibinfo{pages}{779},
  \doiprefix\url{10.1038/s41467-017-00894-w} (\bibinfo{year}{2017}).

\bibitem{Langford2017}
\bibinfo{author}{Langford, N.~K.} \emph{et~al.}
\newblock \bibinfo{journal}{\bibinfo{title}{Experimentally simulating the
  dynamics of quantum light and matter at deep-strong coupling}}.
\newblock {\emph{\JournalTitle{Nature Communications}}}
  \textbf{\bibinfo{volume}{8}}, \bibinfo{pages}{1715},
  \doiprefix\url{10.1038/s41467-017-01061-x} (\bibinfo{year}{2017}).

\bibitem{Chen2022}
\bibinfo{author}{Chen, Y.-H.}, \bibinfo{author}{Miranowicz, A.},
  \bibinfo{author}{Chen, X.}, \bibinfo{author}{Xia, Y.} \&
  \bibinfo{author}{Nori, F.}
\newblock \bibinfo{journal}{\bibinfo{title}{Enhanced-fidelity ultrafast
  geometric quantum computation using strong classical drives}}.
\newblock {\emph{\JournalTitle{Phys. Rev. Appl.}}}
  \textbf{\bibinfo{volume}{18}}, \bibinfo{pages}{064059},
  \doiprefix\url{10.1103/PhysRevApplied.18.064059} (\bibinfo{year}{2022}).

\bibitem{Akbari2025}
\bibinfo{author}{Akbari, K.}, \bibinfo{author}{Nori, F.} \&
  \bibinfo{author}{Hughes, S.}
\newblock \bibinfo{journal}{\bibinfo{title}{Floquet engineering the quantum
  rabi model in the ultrastrong coupling regime}}.
\newblock {\emph{\JournalTitle{Physical Review Letters}}}
  \textbf{\bibinfo{volume}{134}}, \bibinfo{pages}{063602},
  \doiprefix\url{10.1103/physrevlett.134.063602} (\bibinfo{year}{2025}).

\bibitem{Valadares2024}
\bibinfo{author}{Valadares, F.} \emph{et~al.}
\newblock \bibinfo{journal}{\bibinfo{title}{On-demand transposition across
  light-matter interaction regimes in bosonic cqed}}.
\newblock {\emph{\JournalTitle{Nature Communications}}}
  \textbf{\bibinfo{volume}{15}}, \bibinfo{pages}{5816},
  \doiprefix\url{10.1038/s41467-024-50201-7} (\bibinfo{year}{2024}).

\bibitem{Beaudoin2012}
\bibinfo{author}{Beaudoin, F.}, \bibinfo{author}{da~Silva, M.~P.},
  \bibinfo{author}{Dutton, Z.} \& \bibinfo{author}{Blais, A.}
\newblock \bibinfo{journal}{\bibinfo{title}{First-order sidebands in circuit
  qed using qubit frequency modulation}}.
\newblock {\emph{\JournalTitle{Physical Review A}}}
  \textbf{\bibinfo{volume}{86}}, \bibinfo{pages}{022305},
  \doiprefix\url{10.1103/physreva.86.022305} (\bibinfo{year}{2012}).

\bibitem{Strand2013}
\bibinfo{author}{Strand, J.~D.} \emph{et~al.}
\newblock \bibinfo{journal}{\bibinfo{title}{First-order sideband transitions
  with flux-driven asymmetric transmon qubits}}.
\newblock {\emph{\JournalTitle{Physical Review B}}}
  \textbf{\bibinfo{volume}{87}}, \bibinfo{pages}{220505},
  \doiprefix\url{10.1103/physrevb.87.220505} (\bibinfo{year}{2013}).

\bibitem{Wang2019}
\bibinfo{author}{Wang, G.}, \bibinfo{author}{Xiao, R.}, \bibinfo{author}{Shen,
  H.}, \bibinfo{author}{Sun, C.} \& \bibinfo{author}{Xue, K.}
\newblock \bibinfo{journal}{\bibinfo{title}{Simulating anisotropic quantum rabi
  model via frequency modulation}}.
\newblock {\emph{\JournalTitle{Scientific reports}}}
  \textbf{\bibinfo{volume}{9}}, \bibinfo{pages}{4569},
  \doiprefix\url{10.1038/s41598-019-40899-7} (\bibinfo{year}{2019}).

\bibitem{Wallraff2007}
\bibinfo{author}{Wallraff, A.} \emph{et~al.}
\newblock \bibinfo{journal}{\bibinfo{title}{Sideband transitions and two-tone
  spectroscopy of a superconducting qubit strongly coupled to an on-chip
  cavity}}.
\newblock {\emph{\JournalTitle{Phys. Rev. Lett.}}}
  \textbf{\bibinfo{volume}{99}}, \bibinfo{pages}{050501},
  \doiprefix\url{10.1103/PhysRevLett.99.050501} (\bibinfo{year}{2007}).

\bibitem{Leek2009}
\bibinfo{author}{Leek, P.~J.} \emph{et~al.}
\newblock \bibinfo{journal}{\bibinfo{title}{Using sideband transitions for
  two-qubit operations in superconducting circuits}}.
\newblock {\emph{\JournalTitle{Phys. Rev. B}}} \textbf{\bibinfo{volume}{79}},
  \bibinfo{pages}{180511}, \doiprefix\url{10.1103/PhysRevB.79.180511}
  (\bibinfo{year}{2009}).

\bibitem{Novikov2015}
\bibinfo{author}{Novikov, S.} \emph{et~al.}
\newblock \bibinfo{journal}{\bibinfo{title}{Raman coherence in a circuit
  quantum electrodynamics lambda system}}.
\newblock {\emph{\JournalTitle{Nature Physics}}} \textbf{\bibinfo{volume}{12}},
  \bibinfo{pages}{75--79}, \doiprefix\url{10.1038/nphys3537}
  (\bibinfo{year}{2015}).

\bibitem{Lu2017}
\bibinfo{author}{Lu, Y.} \emph{et~al.}
\newblock \bibinfo{journal}{\bibinfo{title}{Universal stabilization of a
  parametrically coupled qubit}}.
\newblock {\emph{\JournalTitle{Phys. Rev. Lett.}}}
  \textbf{\bibinfo{volume}{119}}, \bibinfo{pages}{150502},
  \doiprefix\url{10.1103/PhysRevLett.119.150502} (\bibinfo{year}{2017}).

\bibitem{Dodonov2019}
\bibinfo{author}{Dodonov, A.~V.}, \bibinfo{author}{Napoli, A.} \&
  \bibinfo{author}{Militello, B.}
\newblock \bibinfo{journal}{\bibinfo{title}{Emulation of n -photon
  jaynes-cummings and anti-jaynes-cummings models via parametric modulation of
  a cyclic qutrit}}.
\newblock {\emph{\JournalTitle{Physical Review A}}}
  \textbf{\bibinfo{volume}{99}}, \bibinfo{pages}{033823},
  \doiprefix\url{10.1103/physreva.99.033823} (\bibinfo{year}{2019}).

\bibitem{Wang2017}
\bibinfo{author}{Wang, X.}, \bibinfo{author}{Miranowicz, A.},
  \bibinfo{author}{Li, H.-R.} \& \bibinfo{author}{Nori, F.}
\newblock \bibinfo{journal}{\bibinfo{title}{Observing pure effects of
  counter-rotating terms without ultrastrong coupling: A single photon can
  simultaneously excite two qubits}}.
\newblock {\emph{\JournalTitle{Phys. Rev. A}}} \textbf{\bibinfo{volume}{96}},
  \bibinfo{pages}{063820}, \doiprefix\url{10.1103/PhysRevA.96.063820}
  (\bibinfo{year}{2017}).

\bibitem{Mayero2024}
\bibinfo{author}{Mayero, C.} \& \bibinfo{author}{Omolo, J.~A.}
\newblock \bibinfo{journal}{\bibinfo{title}{Anti-jaynes–cummings interaction
  of a two-level atom with squeezed light: photon statistics, atomic population
  inversion and entropy of entanglement}}.
\newblock {\emph{\JournalTitle{Quantum Information Processing}}}
  \textbf{\bibinfo{volume}{23}}, \doiprefix\url{10.1007/s11128-024-04390-1}
  (\bibinfo{year}{2024}).

\bibitem{Lara2005}
\bibinfo{author}{Rodr\'{i}guez-Lara, B.~M.}, \bibinfo{author}{Moya-Cessa, H.}
  \& \bibinfo{author}{Klimov, A.~B.}
\newblock \bibinfo{journal}{\bibinfo{title}{Combining jaynes-cummings and
  anti-jaynes-cummings dynamics in a trapped-ion system driven by a laser}}.
\newblock {\emph{\JournalTitle{Phys. Rev. A}}} \textbf{\bibinfo{volume}{71}},
  \bibinfo{pages}{023811}, \doiprefix\url{10.1103/PhysRevA.71.023811}
  (\bibinfo{year}{2005}).

\bibitem{Yan2018}
\bibinfo{author}{Yan, F.} \emph{et~al.}
\newblock \bibinfo{journal}{\bibinfo{title}{Tunable coupling scheme for
  implementing high-fidelity two-qubit gates}}.
\newblock {\emph{\JournalTitle{Phys. Rev. Appl.}}}
  \textbf{\bibinfo{volume}{10}}, \bibinfo{pages}{054062},
  \doiprefix\url{10.1103/PhysRevApplied.10.054062} (\bibinfo{year}{2018}).

\bibitem{Xie2014}
\bibinfo{author}{Xie, Q.-T.}, \bibinfo{author}{Cui, S.}, \bibinfo{author}{Cao,
  J.-P.}, \bibinfo{author}{Amico, L.} \& \bibinfo{author}{Fan, H.}
\newblock \bibinfo{journal}{\bibinfo{title}{Anisotropic rabi model}}.
\newblock {\emph{\JournalTitle{Phys. Rev. X}}} \textbf{\bibinfo{volume}{4}},
  \bibinfo{pages}{021046}, \doiprefix\url{10.1103/PhysRevX.4.021046}
  (\bibinfo{year}{2014}).

\bibitem{FriskKockum2019}
\bibinfo{author}{Frisk~Kockum, A.}, \bibinfo{author}{Miranowicz, A.},
  \bibinfo{author}{De~Liberato, S.}, \bibinfo{author}{Savasta, S.} \&
  \bibinfo{author}{Nori, F.}
\newblock \bibinfo{journal}{\bibinfo{title}{Ultrastrong coupling between light
  and matter}}.
\newblock {\emph{\JournalTitle{Nature Reviews Physics}}}
  \textbf{\bibinfo{volume}{1}}, \bibinfo{pages}{19--40},
  \doiprefix\url{10.1038/s42254-018-0006-2} (\bibinfo{year}{2019}).

\bibitem{FornDiaz2019}
\bibinfo{author}{Forn-D\'{i}az, P.}, \bibinfo{author}{Lamata, L.},
  \bibinfo{author}{Rico, E.}, \bibinfo{author}{Kono, J.} \&
  \bibinfo{author}{Solano, E.}
\newblock \bibinfo{journal}{\bibinfo{title}{Ultrastrong coupling regimes of
  light-matter interaction}}.
\newblock {\emph{\JournalTitle{Reviews of Modern Physics}}}
  \textbf{\bibinfo{volume}{91}}, \bibinfo{pages}{025005},
  \doiprefix\url{10.1103/revmodphys.91.025005} (\bibinfo{year}{2019}).

\bibitem{Yu2026}
\bibinfo{author}{Yu, J.-W.} \emph{et~al.}
\newblock \bibinfo{journal}{\bibinfo{title}{Anisotropic rabi model as a noise
  biased qubit}}.
\newblock {\emph{\JournalTitle{arXiv preprint}}}  (\bibinfo{year}{2026}).

\bibitem{Nguyen2019}
\bibinfo{author}{Nguyen, L.~B.} \emph{et~al.}
\newblock \bibinfo{journal}{\bibinfo{title}{High-coherence fluxonium qubit}}.
\newblock {\emph{\JournalTitle{Phys. Rev. X}}} \textbf{\bibinfo{volume}{9}},
  \bibinfo{pages}{041041}, \doiprefix\url{10.1103/PhysRevX.9.041041}
  (\bibinfo{year}{2019}).

\bibitem{Lambert2026}
\bibinfo{author}{Lambert, N.} \emph{et~al.}
\newblock \bibinfo{journal}{\bibinfo{title}{Qutip 5: The quantum toolbox in
  {Python}}}.
\newblock {\emph{\JournalTitle{Physics Reports}}}
  \textbf{\bibinfo{volume}{1153}}, \bibinfo{pages}{1--62},
  \doiprefix\url{10.1016/j.physrep.2025.10.001} (\bibinfo{year}{2026}).

\bibitem{Schuster2005}
\bibinfo{author}{Schuster, D.~I.} \emph{et~al.}
\newblock \bibinfo{journal}{\bibinfo{title}{ac stark shift and dephasing of a
  superconducting qubit strongly coupled to a cavity field}}.
\newblock {\emph{\JournalTitle{Phys. Rev. Lett.}}}
  \textbf{\bibinfo{volume}{94}}, \bibinfo{pages}{123602},
  \doiprefix\url{10.1103/PhysRevLett.94.123602} (\bibinfo{year}{2005}).

\bibitem{Schad2016}
\bibinfo{author}{Schad, P.}, \bibinfo{author}{Shnirman, A.} \&
  \bibinfo{author}{Makhlin, Y.}
\newblock \bibinfo{journal}{\bibinfo{title}{Using majorana spin-$\frac{1}{2}$
  representation for the spin-boson model}}.
\newblock {\emph{\JournalTitle{Phys. Rev. B}}} \textbf{\bibinfo{volume}{93}},
  \bibinfo{pages}{174420}, \doiprefix\url{10.1103/PhysRevB.93.174420}
  (\bibinfo{year}{2016}).

\bibitem{Sete2014}
\bibinfo{author}{Sete, E.~A.}, \bibinfo{author}{Gambetta, J.~M.} \&
  \bibinfo{author}{Korotkov, A.~N.}
\newblock \bibinfo{journal}{\bibinfo{title}{Purcell effect with microwave
  drive: Suppression of qubit relaxation rate}}.
\newblock {\emph{\JournalTitle{Phys. Rev. B}}} \textbf{\bibinfo{volume}{89}},
  \bibinfo{pages}{104516}, \doiprefix\url{10.1103/PhysRevB.89.104516}
  (\bibinfo{year}{2014}).

\bibitem{Ithier2005}
\bibinfo{author}{Ithier, G.} \emph{et~al.}
\newblock \bibinfo{journal}{\bibinfo{title}{Decoherence in a superconducting
  quantum bit circuit}}.
\newblock {\emph{\JournalTitle{Phys. Rev. B}}} \textbf{\bibinfo{volume}{72}},
  \bibinfo{pages}{134519}, \doiprefix\url{10.1103/PhysRevB.72.134519}
  (\bibinfo{year}{2005}).

\bibitem{Krantz2019}
\bibinfo{author}{Krantz, P.} \emph{et~al.}
\newblock \bibinfo{journal}{\bibinfo{title}{A quantum engineer’s guide to
  superconducting qubits}}.
\newblock {\emph{\JournalTitle{Applied Physics Reviews}}}
  \textbf{\bibinfo{volume}{6}}, \doiprefix\url{10.1063/1.5089550}
  (\bibinfo{year}{2019}).

\bibitem{Sank2025}
\bibinfo{author}{Sank, D.}, \bibinfo{author}{Khezri, M.},
  \bibinfo{author}{Isakov, S.} \& \bibinfo{author}{Atalaya, J.}
\newblock \bibinfo{journal}{\bibinfo{title}{Balanced coupling in
  electromagnetic circuits}}.
\newblock {\emph{\JournalTitle{Phys. Rev. Appl.}}}
  \textbf{\bibinfo{volume}{23}}, \bibinfo{pages}{024012},
  \doiprefix\url{10.1103/PhysRevApplied.23.024012} (\bibinfo{year}{2025}).

\bibitem{Chapple2025}
\bibinfo{author}{Chapple, A.~A.}, \bibinfo{author}{Benhayoune-Khadraoui, O.},
  \bibinfo{author}{Richer, S.} \& \bibinfo{author}{Blais, A.}
\newblock \bibinfo{journal}{\bibinfo{title}{Balanced cross-kerr coupling for
  superconducting qubit readout}}.
\newblock {\emph{\JournalTitle{Phys. Rev. Lett.}}}
  \textbf{\bibinfo{volume}{135}}, \bibinfo{pages}{256002},
  \doiprefix\url{10.1103/r4v5-wyyt} (\bibinfo{year}{2025}).

\bibitem{Lin2025}
\bibinfo{author}{Lin, W.-J.} \emph{et~al.}
\newblock \bibinfo{journal}{\bibinfo{title}{Verifying the analogy between
  transversely coupled spin-1/2 systems and inductively-coupled fluxoniums}}.
\newblock {\emph{\JournalTitle{New Journal of Physics}}}
  \textbf{\bibinfo{volume}{27}}, \bibinfo{pages}{033012},
  \doiprefix\url{10.1088/1367-2630/adb77b} (\bibinfo{year}{2025}).

\bibitem{Wallraff2005}
\bibinfo{author}{Wallraff, A.} \emph{et~al.}
\newblock \bibinfo{journal}{\bibinfo{title}{Approaching unit visibility for
  control of a superconducting qubit with dispersive readout}}.
\newblock {\emph{\JournalTitle{Physical Review Letters}}}
  \textbf{\bibinfo{volume}{95}}, \bibinfo{pages}{060501},
  \doiprefix\url{10.1103/physrevlett.95.060501} (\bibinfo{year}{2005}).

\bibitem{Manucharyan2009}
\bibinfo{author}{Manucharyan, V.~E.}, \bibinfo{author}{Koch, J.},
  \bibinfo{author}{Glazman, L.~I.} \& \bibinfo{author}{Devoret, M.~H.}
\newblock \bibinfo{journal}{\bibinfo{title}{Fluxonium: Single cooper-pair
  circuit free of charge offsets}}.
\newblock {\emph{\JournalTitle{Science}}} \textbf{\bibinfo{volume}{326}},
  \bibinfo{pages}{113--116}, \doiprefix\url{10.1126/science.1175552}
  (\bibinfo{year}{2009}).

\bibitem{Groszkowski2021}
\bibinfo{author}{Groszkowski, P.} \& \bibinfo{author}{Koch, J.}
\newblock \bibinfo{journal}{\bibinfo{title}{Scqubits: a {P}ython package for
  superconducting qubits}}.
\newblock {\emph{\JournalTitle{{Quantum}}}} \textbf{\bibinfo{volume}{5}},
  \bibinfo{pages}{583}, \doiprefix\url{10.22331/q-2021-11-17-583}
  (\bibinfo{year}{2021}).

\bibitem{Hazard2019}
\bibinfo{author}{Hazard, T.~M.} \emph{et~al.}
\newblock \bibinfo{journal}{\bibinfo{title}{Nanowire superinductance fluxonium
  qubit}}.
\newblock {\emph{\JournalTitle{Phys. Rev. Lett.}}}
  \textbf{\bibinfo{volume}{122}}, \bibinfo{pages}{010504},
  \doiprefix\url{10.1103/PhysRevLett.122.010504} (\bibinfo{year}{2019}).

\bibitem{Marxer2023}
\bibinfo{author}{Marxer, F.} \emph{et~al.}
\newblock \bibinfo{journal}{\bibinfo{title}{Long-distance transmon coupler with
  cz-gate fidelity above $99.8\rm{\%}$}}.
\newblock {\emph{\JournalTitle{PRX Quantum}}} \textbf{\bibinfo{volume}{4}},
  \bibinfo{pages}{010314}, \doiprefix\url{10.1103/PRXQuantum.4.010314}
  (\bibinfo{year}{2023}).

\bibitem{Chen2016}
\bibinfo{author}{Chen, Z.} \emph{et~al.}
\newblock \bibinfo{journal}{\bibinfo{title}{Measuring and suppressing quantum
  state leakage in a superconducting qubit}}.
\newblock {\emph{\JournalTitle{Phys. Rev. Lett.}}}
  \textbf{\bibinfo{volume}{116}}, \bibinfo{pages}{020501},
  \doiprefix\url{10.1103/PhysRevLett.116.020501} (\bibinfo{year}{2016}).

\bibitem{Sung2021}
\bibinfo{author}{Sung, Y.} \emph{et~al.}
\newblock \bibinfo{journal}{\bibinfo{title}{Realization of high-fidelity cz and
  $zz$-free iswap gates with a tunable coupler}}.
\newblock {\emph{\JournalTitle{Phys. Rev. X}}} \textbf{\bibinfo{volume}{11}},
  \bibinfo{pages}{021058}, \doiprefix\url{10.1103/PhysRevX.11.021058}
  (\bibinfo{year}{2021}).

\bibitem{Pozar2021}
\bibinfo{author}{Pozar, D.~M.}
\newblock \emph{\bibinfo{title}{Microwave engineering: theory and techniques}}
  (\bibinfo{publisher}{John wiley \& sons}, \bibinfo{year}{2021}).

\bibitem{Loewdin1950}
\bibinfo{author}{L{\"o}wdin, P.-O.}
\newblock \bibinfo{journal}{\bibinfo{title}{On the non-orthogonality problem
  connected with the use of atomic wave functions in the theory of molecules
  and crystals}}.
\newblock {\emph{\JournalTitle{The Journal of Chemical Physics}}}
  \textbf{\bibinfo{volume}{18}}, \bibinfo{pages}{365--375}
  (\bibinfo{year}{1950}).

\bibitem{Loewdin1962}
\bibinfo{author}{L{\"o}wdin, P.-O.}
\newblock \bibinfo{journal}{\bibinfo{title}{Studies in perturbation theory. iv.
  solution of eigenvalue problem by projection operator formalism}}.
\newblock {\emph{\JournalTitle{Journal of Mathematical Physics}}}
  \textbf{\bibinfo{volume}{3}}, \bibinfo{pages}{969--982}
  (\bibinfo{year}{1962}).

\bibitem{Winkler2003}
\bibinfo{author}{Winkler, R.}
\newblock \emph{\bibinfo{title}{Spin-Orbit Coupling Effects in Two-Dimensional
  Electron and Hole Systems}} (\bibinfo{publisher}{Springer Berlin,
  Heidelberg}, \bibinfo{year}{2003}).
\newblock \bibinfo{note}{See Appendix B: L\"{o}wdin Partitioning}.

\end{thebibliography}

\end{document}